\RequirePackage{fix-cm}
\documentclass[smallextended]{svjour3}       
\smartqed  
\usepackage[labelfont=bf, labelsep=space]{caption}
\usepackage{graphicx}
\usepackage{fontawesome}
\usepackage{array}
\usepackage{lscape}
\usepackage{multirow}
\usepackage{graphicx}
\usepackage{rotating}
\usepackage{makecell}
\usepackage{booktabs, tabularx}
\usepackage{graphicx}
\usepackage{ltablex}
\usepackage{amsmath}
\usepackage[tableposition=t]{caption}
\newcolumntype{L}{>{\raggedright\arraybackslash}X}
\usepackage{comment}
\usepackage{afterpage}
\usepackage{enumitem}%
\usepackage{multicol}
\setlist{nosep, noitemsep}
\usepackage{enumitem}
\usepackage{comment}
\usepackage{booktabs} 
\usepackage{pdfpages}
\usepackage{soul}
\usepackage{pdflscape}
\usepackage{subfig}
\usepackage{fontawesome}
\usepackage{tabularx}
\usepackage{makecell}
\usepackage{tcolorbox}
\usepackage{dirtytalk}
\usepackage{float}
\usepackage{multirow}
\usepackage{supertabular}
\usepackage{longtable}
\usepackage{svg}

\usepackage[round]{natbib}

\usepackage[hidelinks]{hyperref}
\hypersetup{
  colorlinks   = true, 
  urlcolor     = blue, 
  linkcolor    = blue, 
  citecolor   = blue 
}

\usepackage{etoolbox}

\makeatletter

\let\oldciteauthor\citeauthor

\def\citeauthor#1{{\NoHyper\oldciteauthor{#1}}}

\patchcmd{\NAT@citex}
  {\@citea\NAT@hyper@{%
     \NAT@nmfmt{\NAT@nm}%
     \hyper@natlinkbreak{\NAT@aysep\NAT@spacechar}{\@citeb\@extra@b@citeb}%
     \NAT@date}}
  {\@citea\NAT@nmfmt{\NAT@nm}%
   \NAT@aysep\NAT@spacechar\NAT@hyper@{\NAT@date}}{}{}

\patchcmd{\NAT@citex}
  {\@citea\NAT@hyper@{%
     \NAT@nmfmt{\NAT@nm}%
     \hyper@natlinkbreak{\NAT@spacechar\NAT@@open\if*#1*\else#1\NAT@spacechar\fi}%
       {\@citeb\@extra@b@citeb}%
     \NAT@date}}
  {\@citea\NAT@nmfmt{\NAT@nm}%
   \NAT@spacechar\NAT@@open\if*#1*\else#1\NAT@spacechar\fi\NAT@hyper@{\NAT@date}}
  {}{}

\makeatother

\begin{document}

\title{Collaborative Application Security Testing \\ for DevSecOps}
\subtitle{An Empirical Analysis of Challenges, Best Practices \\ and Tool Support}


\author{Roshan Namal Rajapakse \and
        Mansooreh Zahedi \and Muhammad Ali Babar
}


\institute{Roshan Namal Rajapakse, Muhammad Ali Babar \at
              \textsuperscript{1}Centre for Research on Engineering Software Technologies, School of Computer Science, \\ The University of Adelaide, Australia. \textsuperscript{2}Cyber Security Cooperative Research Centre, Australia.
              \email{\underline{roshan.rajapakse@adelaide.edu.au}, ali.babar@adelaide.edu.au}           
           \and
           Mansooreh Zahedi \at
              School of Computing and Information Systems, The University of Melbourne, Australia.\\
              \email{mansooreh.zahedi@unimelb.edu.au}
}

\date{Received: date / Accepted: date}

\maketitle

\begin{abstract}
\textit{DevSecOps} is a software development paradigm that places a high emphasis on the \textit{culture of collaboration} between developers (\textit{Dev}), security (\textit{Sec}) and operations (\textit{Ops}) teams to deliver secure software continuously and rapidly. Adopting this paradigm, therefore, requires an understanding of the challenges each team faces when doing collaborative work and strategies for overcoming them. However, collaborative aspects related to these teams have received very little empirical attention in the literature. Here we present a study focusing on a key security activity, Application Security Testing (AST), in which practitioners find it difficult to collaborate in a DevSecOps environment. We also show how a systematically selected set of webinars, technical talks and panel discussions can be used as a data source to qualitatively analyse practitioner views on the most recent trends and emerging solutions of a highly evolving field. We find that the lack of features that facilitate a collaborative workflow built into the AST tools themselves is a key tool-related challenge in DevSecOps. In addition, the lack of clarity related to role definitions, shared goals, and ownership also hinders Collaborative AST (CoAST). We also captured a range of best practices for collaboration (e.g., \textit{Shift-left security}), emerging communication methods (e.g., \textit{ChatOps}), and new team structures (e.g., \textit{hybrid teams}) for CoAST. Finally, our study identified several requirements for new tool features and specific gap areas for future research to provide better support for CoAST in DevSecOps.
\end{abstract}

\section{Introduction}
As the software development industry strives for greater speeds in deploying software outputs, the Development and Operations (\textit{DevOps}) paradigm is becoming increasingly popular. The key idea of DevOps is to remove boundaries (or \textit{silos}) between the development and operations teams to enable better collaboration, broaden skillsets and share responsibilities among team members \citep{jabbari2016devops}. While collaboration is critical for all aspects of DevOps, many studies \citep{mohan2016secdevops, mohan2018bp, sanchez2020security, lwakatare2015dimensions, mao2020preliminary, Rajapakse_Zahedi_Babar_Shen_2022a, akbar2022toward} have particularly highlighted the importance of having a strong \textit{culture of collaboration} to produce secure outputs in a rapid deployment environment. This requirement for integrating security in DevOps has led to an emerging paradigm named \textit{DevSecOps} (i.e., Development (Dev), Security (Sec) and Operations (Ops). Similar to DevOps, the paradigm shift of DevSecOps is related to enabling the security team better collaborate with developers and operators by removing the functional boundaries between them \citep{Rajapakse_Zahedi_Babar_Shen_2022a, lwakatare2015dimensions}. 

In DevSecOps, one specific activity that is considered highly collaborative is Application Security Testing (AST), which purports to identify and rectify software security vulnerabilities \citep{Gartner}. Traditionally, AST was performed by the security team at the end of the software development process, typically as a time-consuming quality checking activity \citep{Rajapakse_Zahedi_Babar_Shen_2022a}. However, with the need to rapidly and continuously deploy software outputs in DevSecOps, AST activities are required to be done as early as possible in order to receive early feedback (i.e., \textit{shift-left security} \citep{rajapakse2021empirical}). As a result, software developers are now required to engage in AST-related tasks while conducting development activities. However, as developers typically lack security knowledge and the required AST expertise, they are expected to frequently collaborate with the security team to conduct the relevant security tasks.

\begin{figure*} [t]
    \centering
    \includegraphics[scale=.22]{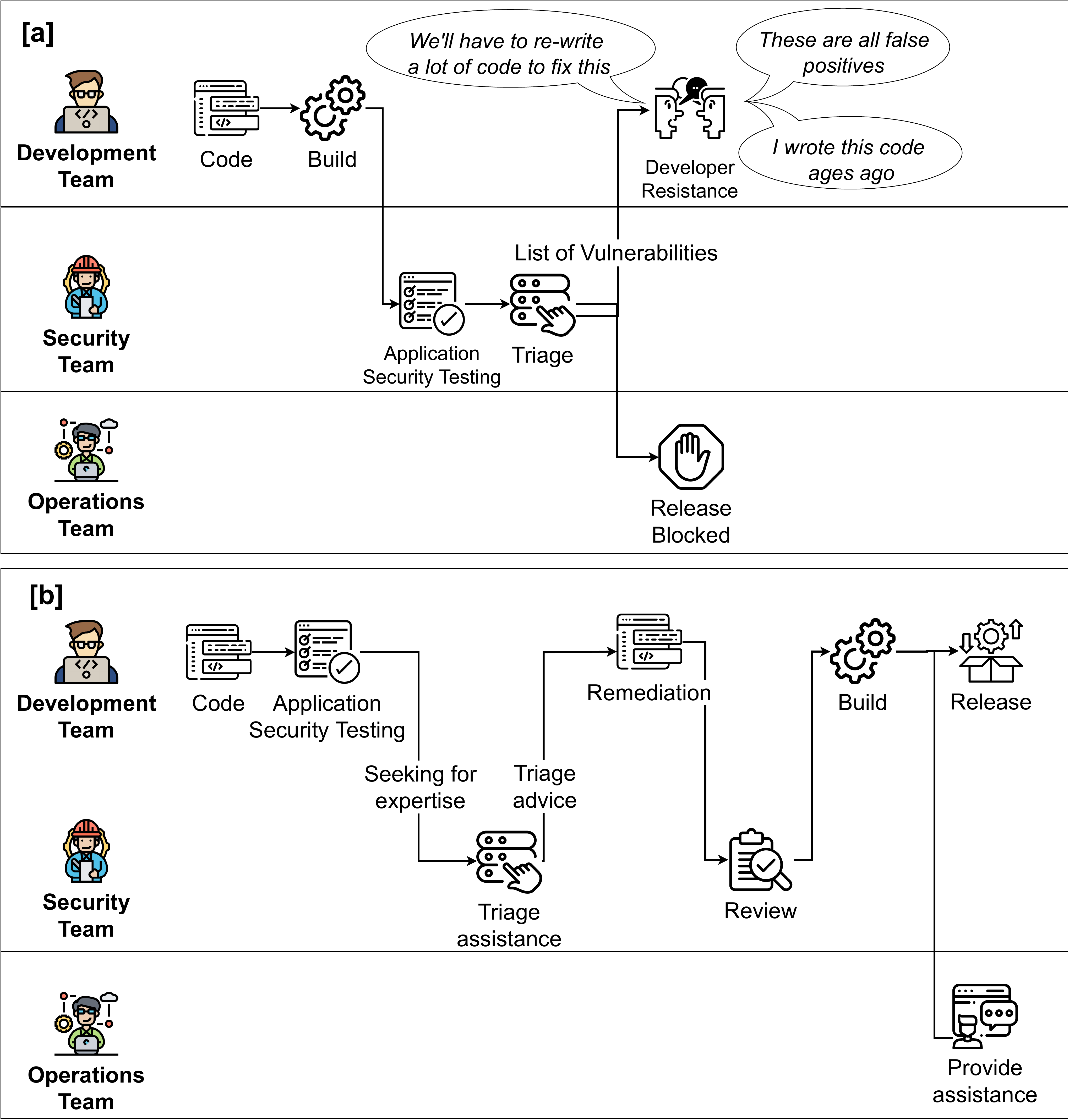}
    \caption{Two specific practical examples (a,b) for collaboration difficulties practitioners face in performing AST in a DevSecOps environment}
    \label{fig:case1}
\end{figure*}

However, given that collaboration in software development is known to be a challenging undertaking, practitioners \footnote{We use the term 'practitioners' to collectively refer to all roles or stakeholders engaged in a DevSecOps environment.} have reported considerable difficulties in performing \textit{Collaborative Application Security Testing} (CoAST) in a DevSecOps environment \citep{mohan2018bp, tahaei2019survey, mao2020preliminary, Rajapakse_Zahedi_Babar_Shen_2022a}. In fact, \textit{poor collaborative organizational culture} \citep{khan2020multicriteria} is frequently reported as a challenging factor in DevSecOps environments \citep{hussain2017emerging, diel2016communication, bass2015devops, hemon2020agile, calefato2019agile}. 

To offer an overview of these difficulties to the reader, we present two specific examples of CoAST challenges, as pointed out by practitioners [W23] and the related studies \citep{tahaei2019survey, rajapakse2021empirical}. In the first case (Fig. \ref{fig:case1}[a]), AST is conducted by the security team working with separate sets of tools and processes. The discovered vulnerabilities are communicated to developers (at a later stage of the software development life cycle) and the operations team via communications channels such as emails or ticketing systems. Such a workflow has frequently resulted in collaboration problems such as a lack of shared understanding of the overall AST or DevOps goals and limitations related to leveraging the expertise of other teams in a timely manner \citep{tahaei2019survey}. Studies also state that in this sort of workflow, there is no clarity related to communicating security issues across teams, such as the specific types of technical information that need to be shared or tracked \citep{tahaei2019survey, Rajapakse_Zahedi_Babar_Shen_2022a}. As a result of these difficulties, such a working environment typically causes developer resistance to AST [W23].

As recommended by DevSecOps, AST is done by the development team in the second case (Fig. \ref{fig:case1}[b]). However, limitations related to AST tools and platforms hinder collaboration in this workflow. Here, a software developer performs AST (e.g., static code analysis) prior to the build stage. Yet, as most developers lack in-depth security expertise, they need to seek the assistance of the security team several times during the workflow (e.g., assistance for vulnerability triage and review of the remedial measures taken to approve build and subsequent release). However, software developers report frustrations with AST tools that do not adequately support such frequent collaboration in this context \citep{tahaei2019survey}. Many difficulties related to sharing technical details (e.g., security scan results, code snippets that require modifications) with the security team, lack of convenient workflows into bug trackers, and unavailability of teamwork features have been highlighted in this regard \citep{tahaei2019survey, gibler_2022}. Based on the above cases, it is clear that this area needs more focus to support better collaborative work.

Considering the significance of this area for a modern software development setting and the lack of in-depth systematic work in these specific challenges, we decided to empirically explore different aspects of CoAST in DevSecOps. We first aimed to identify and classify the challenges faced by practitioners in this area. The lack of adequate awareness of these challenges may be one reason for the reported issues. We also aimed to identify the best practices for collaboration recommended by practitioners to facilitate CoAST. Finally, to aid the advancements in tool development in this area, we planned to synthesise the existing AST tool landscape for CoAST and then analyse the available solutions against the identified challenges. This effort would assist us in identifying what needs to be done to improve the existing AST tools or establish requirements for new tool developments.

In this study, we report our novel use of 48 systematically selected webinars, technical talks and panel discussions \footnote{We shorten these sources to the term 'webinars' in the rest of the paper.} as a data source to achieve the above aims. As a highly evolving field, we recognised that capturing the emerging technologies and latest trends in the industry would be essential in this study. Therefore, we used webinars where industry experts and experienced practitioners discussed the current state of the field. Our aim was to utilise this data source, rich with practice-based information, to capture the most up-to-date and relevant data for our research questions (RQs).

\begin{tcolorbox}[left=1pt, top=1pt, right=1pt, bottom=1pt]
Our \textbf{contributions} to the current body of knowledge on CoAST targeting DevSecOps are as follows. 
\begin{itemize}
\item To the best of our knowledge, this is the first in-depth study that systematically analyses practitioners' perspectives on different aspects of \textit{CoAST} in DevSecOps.
\item We present a body of knowledge on CoAST challenges faced by practitioners, best practices for collaboration and tool support targeting CoAST. Our findings can inform the research community about the reported challenges to devise future studies and act as guidelines for practitioners to adopt CoAST.
\item We present specific feature requirements for new tool developments to aid tool producers in this area.
\item We also highlight the methodological benefits that can be gained from utilising our selected type of data and method for software engineering research on emerging areas.
\end{itemize}
\end{tcolorbox}

\section{Background and Related work}

In this section, we present and discuss the key concepts that underpin the DevSecOps paradigm and the related studies to contextualise our research reported in this paper.

\subsection{Security in DevOps}
Security in DevOps (i.e., DevSecOps) refers to the integration of security principles and controls (e.g., processes, tools) into the DevOps workflow \citep{myrbakken2017devsecops, mohan2016secdevops}. Here, the main aim is to assess and ensure the security of the software outputs while deploying continuously and rapidly \citep{Rajapakse_Zahedi_Babar_Shen_2022a}. The literature on DevSecOps has been reviewed in different studies, which provide the definitions \citep{mohan2016secdevops, myrbakken2017devsecops}, metrics \citep{prates2019devsecops}, culture \citep{sanchez2020security}, adoption challenges and solutions \citep{Rajapakse_Zahedi_Babar_Shen_2022a}, enablers and recommended practices of DevSecOps \citep{mao2020preliminary}. The recommendations related to integrating security in these studies can be largely synthesised into three pillars: \textit{people}-related or socio-technical changes, \textit{process} improvements and \textit{technological} (e.g., tools, infrastructure) adoptions \citep{Rajapakse_Zahedi_Babar_Shen_2022a}. Regarding the first pillar (i.e., people), the most widely reported recommendation is promoting collaboration among the development, operations, and security teams of an organisation \citep{mohan2016secdevops, sanchez2020security, mao2020preliminary, Rajapakse_Zahedi_Babar_Shen_2022a}. For example, increased collaboration with the security team is considered vital to move security early into the DevOps workflow (i.e., shift-left security) \citep{Rajapakse_Zahedi_Babar_Shen_2022a, mao2020preliminary}. In this scenario, developers are required to engage in security assessment tasks while engaging in the development activities \citep{Rajapakse_Zahedi_Babar_Shen_2022a}. 

To implement the above recommendations, a commonly accepted key enabler is adopting suitable security tools in the DevSecOps workflow \citep{myrbakken2017devsecops, rafi2020prioritization, Rajapakse_Zahedi_Babar_Shen_2022a, mao2020preliminary}. For instance, developers should be provided with the appropriate AST tools that they can use while developing software in rapid deployment cycles. However, many studies highlight that security tool-related limitations are a significant challenge in this domain \citep{myrbakken2017devsecops, rafi2020prioritization, rajapakse2021empirical, Rajapakse_Zahedi_Babar_Shen_2022a}. Myrbakken and Colomo-Palacios \citep {myrbakken2017devsecops} present work that describes the difficulties faced by developers in producing secure code in DevSecOps while using security tools. As the major contributing factor to this issue, some studies highlight that traditional AST tools might not have been developed targeting principles and practices of modern software development paradigms \citep{Rajapakse_Zahedi_Babar_Shen_2022a}. Rather most of the traditional tools focus on the needs of security teams, which usually work during the later stages (e.g., testing or verification) of traditional software development methodologies (i.e., waterfall method). That is why there are many drawbacks related to traditional AST tools in the context of DevSecOps \citep{rajapakse2021empirical, Rajapakse_Zahedi_Babar_Shen_2022a}. 

\subsection{Application security testing in DevOps}
Application security testing (AST) is the process that plays a major role in 
producing secure software by identifying security weaknesses and vulnerabilities in source code \citep{Gartner}. While AST started as a manual activity, it is nowadays enabled by different \textit{types} of AST tools available in the industry \citep{scanlon_2018}. For example, Static Application Security Testing (SAST) \citep{Yang2019} and Dynamic Application Security Testing (DAST) tools \citep{rangnau2020continuous} are popular among practitioners. In addition, there are tools such as Interactive Application Security Testing (IAST) and Software Composition Analysis (SCA), which are emerging in the industry \citep{rajapakse2021empirical}. Organisations typically employ AST programs that combine several of these AST tool types and the related processes to identify and address security vulnerabilities in the software \citep{bird2014survey}. However, several studies point out many drawbacks of such AST tools when used in rapid deployment environments such as DevOps \citep{Rajapakse_Zahedi_Babar_Shen_2022a}.

Mao et al., \citep{mao2020preliminary} presented the impact of DevOps on software security, practitioners' perceptions, and practices associated with DevSecOps. They identified that the lack of \textit{mature} tools suitable for DevSecOps is a significant implementation challenge in this domain. Here, the authors state that in DevOps, the security functionality has to be available in tools that work on the right platforms (i.e., developer work environment) \citep{myrbakken2017devsecops}. However, the issue is that there are a limited number of mature tools that are \textit{efficient enough to keep up with DevOps' speed} \citep{myrbakken2017devsecops}

Studies also point towards the drawbacks of specific AST tools in this context. For example, Rajapakse et al. \citep{Rajapakse_Zahedi_Babar_Shen_2022a} report that static (i.e., SAST) and dynamic analysis (i.e., DAST)  tools, in particular, are difficult to be integrated into the rapid deployment cycles of DevOps. Regarding SAST, high numbers of false positives \citep{tomas2019empirical, johnson2013don, zampetti2017open, croft2021empirical}, the lengthy code scanning time, and high resource consumption \citep{soenen2018insights} are some of the frequently reported drawbacks. Similarly, in DAST, the requirement of manual intervention, the lengthy time to run the tool (as the code needs to be built and configured) \citep{kupsch2017continuous, brady2020docker}, and the limited scope of testing scenarios are some of the limitations of this type of tool highlighted in the literature \citep{siewruk2019security}. As a result of these limitations, there appears to be a tendency among practitioners not to utilise these tools in their DevOps workflows despite the potential benefits of ensuring security.

Due to the above-mentioned issues, practitioners highlight the need to move towards new generations of AST tools that target DevOps requirements \citep{rajapakse2021empirical}. As a result, tools producers are now promoting new toolsets and technologies to satisfy these emerging requirements (\citep {451Research2018, Synopsys2020, Outpost24}). However, many of these new tools' capabilities have not been empirically assessed against DevSecOps requirements in the literature \citep{Rajapakse_Zahedi_Babar_Shen_2022a}. 

We have also noted that many of the early review studies in this domain (e.g., \citep{mohan2016secdevops, prates2019devsecops, sanchez2020security}) have included small numbers of peer-reviewed papers. Accordingly, the findings of these studies are limited to the few resources captured. As a heavily industry-centric field, there is also a need for DevSecOps studies that have used industry resources or practitioner data. Such resources are important to uncover the \textit{state of practice} in the industry \citep{garousi2016need}. However, based on the available literature, it is apparent that there is a limited number of studies that have utilised industry-centric data. Therefore, in this study, we used popular forms of engagement among practitioners: webinars, technical talks and panel discussions. The recordings of these sources are freely available on YouTube. We found that our sources, which are conducted and attended by large numbers of practitioners, are rich with up-to-date, industry-oriented (i.e., practice-based) information. For example, the webinars included speakers from the industry discussing actual ground-level problems and tool producers presenting emerging solutions in the domain.

The closest study to this paper is our previous work which used webinars to empirically investigate challenges faced by practitioners when integrating security tools into a DevOps workflow \citep{rajapakse2021empirical}. In this previous work, we presented a set of recommended security tools and services for DevOps and then provided guidelines to use them in a DevOps workflow. In contrast, this study contributes to the literature by focusing on a specific requirement, \textit{cross-functional team collaboration} in a DevSecOps environment. We also limit the focus to Application Security Testing to investigate the challenges, practices and tool support for collaboration. In order to deeply investigate this specific focus, we further increased the number of webinar sources in this study. Here, we do not limit our study to a single source (i.e., YouTube channel) as done in our previous work \citep{rajapakse2021empirical}, but expand into other popular channels. Due to the above reasons, there is only an overlap of 16 webinars with our previous study. 

\subsection{Existing research on Collaborative Software Engineering}
Software engineering is inherently a collaborative process, as a software system is rarely built by one person. Therefore, any software development activity typically involves a number of stakeholders working as a team using a process of \textit{Collaborative Software Engineering} (CoSE) \citep{Mistrik20101}. There is also a growing body of literature dealing with collaborative practices, tools, and operational research related to organisational structures and processes. However, several studies state that many challenges are still present related to the above-noted areas and improving the state-of-the-art \citep{Mistrik20101, treude2009empirical}.

With regard to the existing research in CoSE, most studies have been done in the areas of global software engineering \citep{richardson2010global, zahedi2016systematic}, agile software development \citep{robinson2010collaboration}, model-driven software engineering \citep{franzago2017collaborative, david2021collaborative}, requirements engineering \citep{damian2010requirements}, open source software development \citep{scacchi2010collaboration}, and collaboration tools \citep{lanubile2010collaboration, hattori2010syde}.  Based on these studies, socio-technical concerns are closely interrelated with collaborative aspects, especially in the initial phases of the development cycle \citep{treude2009empirical}. However, the challenges associated with collaboration become more difficult when the team members are geographically distributed, which introduces various spatial, temporal, and socio-cultural issues \citep{treude2009empirical, whitehead2010collaborative}. 

Artifacts in software projects play a crucial role in CoSE \citep{treude2009empirical}. This includes creating and negotiating shared meaning using project artifacts that contain models describing the final outcomes \citep{whitehead2010collaborative}. Here, model-driven collaboration using the shared models has received particular attention in the literature \citep{franzago2017collaborative, david2021collaborative}. Researchers have named this area \textit{Collaborative Model-Driven Software Engineering}, which includes techniques that enable different team members of software projects to manage, collaborate, and be aware of each other's work on a set of shared models \citep{franzago2017collaborative}. Similarly, we address collaborative aspects in this work but with a specific focus on AST as it relates to rapid deployment paradigms such as DevSecOps (i.e., \textit{CoAST}).

Studies also state that proper tool support (e.g., Collaboration platforms) is required for maintaining adequate collaboration among software teams \citep{treude2009empirical, sarma2010continuous}. However, a key problem in this domain is the lack of established metrics for quantitatively assessing collaborative aspects in software projects \citep{whitehead2010collaborative}. This results in difficulties in evaluating tool adoption decisions or whether a new tool has made any specific improvements. This is a specific challenge that we discuss in relation to CoAST in our study.

Lastly, researchers state that depending on the development context, there are numerous dimensions of collaboration, which ultimately result in different forms of collaboration requirements \citep{Mistrik20101}. CoSE studies further conclude that tools and techniques that work in one context may not be successful in another setting \citep{Mistrik20101}. Accordingly, new studies are required to assess collaborative aspects related to new software development paradigms which are centred around collaboration (e.g., DevOps and DevSecOps). Our study intends to address this gap area with a specific focus on AST.

\section{Methodology}

\begin{figure*}
    \centering
    \includegraphics[width=\columnwidth]{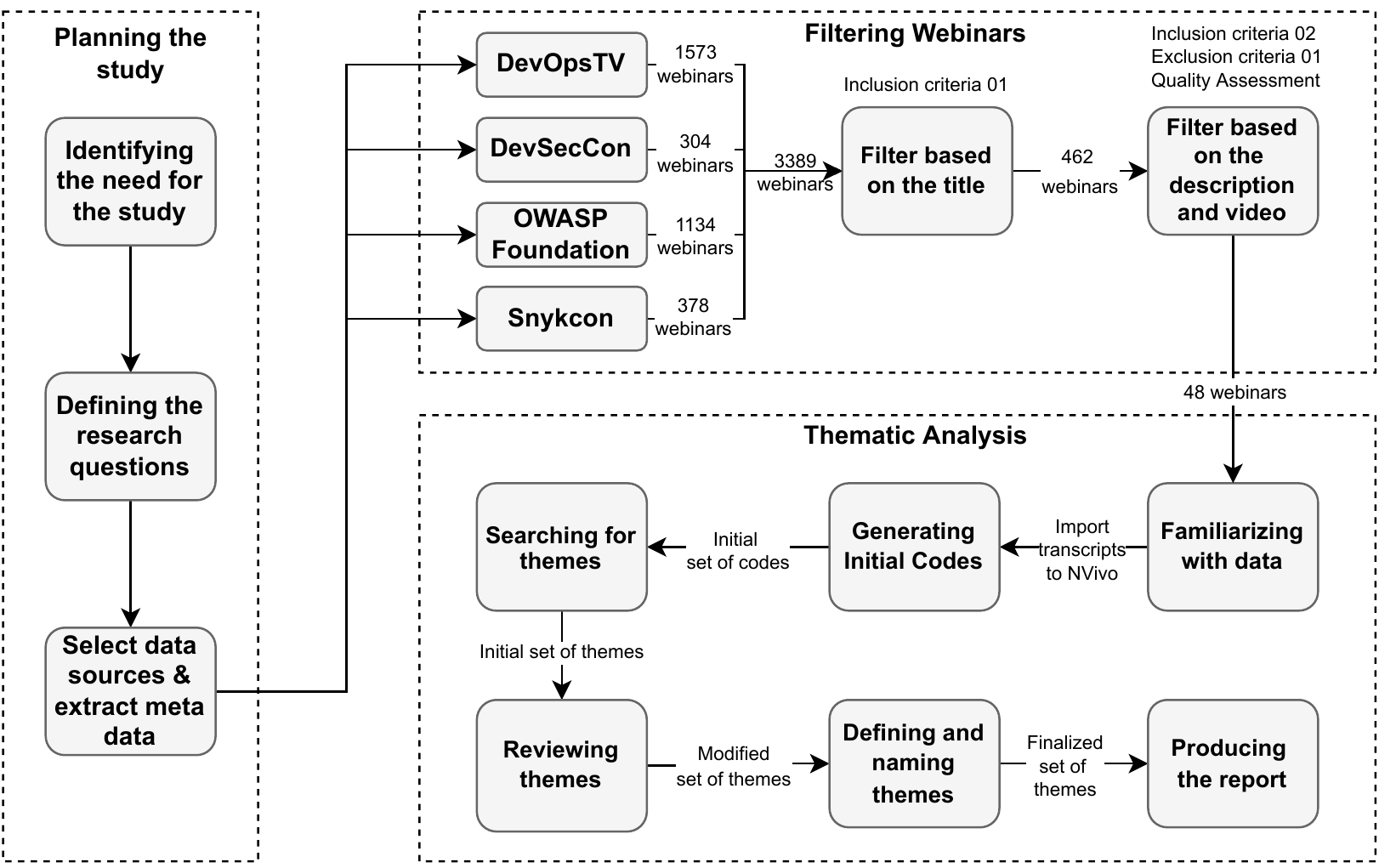}
    \caption{The overview of the method followed in this study}
    \label{fig:method}
\end{figure*}

This section presents the methodology we followed in our study (Fig. \ref{fig:method}).

\subsection{Research questions (RQs)}

To achieve our objectives and guide the analysis, we formed the following RQs.

\begin{tcolorbox}[left=4pt, top=2pt, right=4pt, bottom=2pt]
\textbf{RQ1}: What are the specific challenges related to performing CoAST in a DevSecOps environment?

RQ1 aims to thematically classify the challenges developers, security, and operations engineers face when performing AST collaboratively in DevSecOps.

\textbf{RQ2}: What are the best practices practitioners recommend to perform CoAST in DevSecOps?

RQ2 aims to synthesise and report the specific best practices for collaboration recommended by practitioners to perform AST in DevSecOps.

\textbf{RQ3}: What are the tool types and capabilities that support collaborative work for AST in DevSecOps?

RQ3 aims to identify the currently available or emerging tool types and capabilities that support CoAST. These results would then be analysed against the findings of RQ1 to specify the improvements for currently available products and to establish requirements for new tools.

\end{tcolorbox}

\begin{table*}[b!]
\footnotesize
\centering
\caption{Sample record for meta extracted from a webinar}

\begin{tabular}
{ | m{.12\textwidth} | m{.8\textwidth}| } 
\toprule 

\textbf{ID} & \textbf{Data} \\ 
\hline 
 
Author & DevOpsTV  \\
\hline 
Title & Security in CI/CD Pipelines: Tips for DevOps Engineers \\
\hline
ULR & \url{https://www.youtube.com/watch?v=S7TfXEyhLck} \\
\hline
Views/Likes & 3304/55 \\
 \hline
Upload date & 2019/09/18 \\
 \hline 
Tags & devops.com, devops, devsecops, devopstv, CI/CD, Wallarm, API, cloud-native, K8 \\
 \hline 

Description (Shortened) & While DevOps is becoming a new norm for most of the companies, security is typically still behind. [..] Will cover incorporating security into existing CI/CD pipelines and tools DevOps professionals need to know to implement the automation and adhere to secure coding practices. Join Stepan Ilyin, Chief Product Officer at Wallarm for an engaging conversation where you’ll learn: Methodologies and tooling for dynamic and static security testing [..].\\
  
\bottomrule

\end{tabular}
\label{table:sample}
\end{table*}

\subsection{Source selection and methodological benefits}

As a fast-moving field, many tools, frameworks and best practices are introduced frequently in the DevOps and DevSecOps industry. Therefore, we recognised the need to capture the most recent developments or trends in this industry to adequately answer our RQs. Furthermore, based on our previous studies \citep{rajapakse2021empirical, Rajapakse_Zahedi_Babar_Shen_2022a}, we also recognised that some of the content reported in the peer-reviewed literature in this domain is a considerable time behind compared with the solutions available in the industry. Accordingly, we sought to find a data source that contains the most up-to-date information about the emerging technologies and the state of practice of DevSecOps. We achieved this objective by selecting webinars as our data source. Below, we describe the \textbf{methodological benefits} received from using this data source for our study.

\begin{itemize}
\item Our aim was to use a highly systematic selection and quality assessment (QA) process on a 1st or 2nd tier grey-literature source \citep{garousi2019guidelines}. Selecting webinars (technical talks and panel discussions) from a set of reputed Youtube channels enabled us to employ such a process. The meta data of webinars contained components similar to a peer-reviewed paper (i.e., title, abstract (i.e., description), keywords (i.e., tags) and content (i.e., the video)) that we could use to follow a systematic selection process (Table \ref{table:sample}). We were also able to use these components to adequately perform a QA stage to remove webinars with a heavy commercial focus.
\item The selected webinars were from reputed and highly viewed channels on Youtube for such content. Further, they were mostly conducted by industry experts (e.g., independent consultants, tool producers, product engineers, and chief technical/executive officers) in this domain. Therefore, most speakers had extensive experience and were quite knowledgeable about the topic being discussed. This is a benefit compared to conducting interview studies, where it would be difficult to attract a large number of experienced industry experts.

\item Some webinars presented tool demonstrations from leading tool producers. Such videos enabled us to capture to latest tools and emerging technologies in this domain. Based on these demos, we were also able to assess for ourselves whether these technologies were adequate to address the current challenges reported by practitioners. 

\item Webinars offered numerous and, at times, diverse views on our topic of interest (e.g., panel discussions). We benefited from such discussions to capture contrasting practitioner viewpoints regarding proposed solutions.

\item It is already established that software practitioners typically do not have the time, access or expertise to read peer-reviewed academic content \citep{garousi2020benefitting}. As a result, practitioners are likely to turn to grey resources. We benefited from this because most webinars contained a question-and-answer session where practitioners (e.g., attendees or viewers online) discussed their ideas, experiences and challenges in detail with the speakers. 
\end{itemize}

\subsection{Webinar selection strategy}

In order to select the webinars most relevant to our study, we first used the search function of YouTube to select popular channels that post DevSecOps-related content. To select a YouTube channel as the source, we considered whether it consisted of a) up-to-date, b) regularly posted c) vendor-neutral content. We also considered the popularity of the channel based on the number of subscribers. At the end of this search, we selected four channels for our study (Table \ref{table:webinars}).

\begin{table*}[b!]
\footnotesize
\centering
\caption{Sources for the webinars, technical talks and panel discussions}
\begin{tabular}{ | m{.11\columnwidth} | m{.6\columnwidth} | m{.2\columnwidth} | } 
\toprule 
\textbf{Name} & \textbf{Description \& (the number of subscribers*5/22)} & \textbf{Link} \\ 
 \hline 
 DevOpsTV & The Youtube channel of \url{https://www.DevOps.com}: One of the largest collections of DevOps Webinars online (12,800) & \url{https://www.youtube.com/c/Devopsdotcom} \\ 
 \hline 
 DevSecCon & the Youtube channel of a global DevSecOps community: \url{https://www.devseccon.com/} (2,020) &  \url{https://www.youtube.com/c/DevSecCon} \\ 
  \hline
 OWASP Foundation & Technical talks from OWASP AppSec conferences (53,800) & \url{https://www.youtube.com/c/OWASPGLOBAL} \\ 
 \hline
 SnykCon & Technical talks from SnykCon: A popular developer conference which emphasizes developer-centric security (3,560) & \url{https://www.youtube.com/c/Snyksec/videos}
 \\  \bottomrule
\end{tabular}
\label{table:webinars}
\end{table*}

The selected channels contained a large number of videos in various sub-domains and issues. Therefore, our first task was to select webinars related our research questions. We followed the below noted steps for this purpose.

\begin{itemize}
    \item  Firstly, we utilised a python API \citep{pypi} to extract the relevant metadata (e.g., Title, Link, View count, Date, Tags, and Description) from all the available videos on the selected channels on the 28th of January 2022. A combined 3389 videos were available across the four channels.
    \item  We then filtered the webinars based on the title to select webinars with an Application Security Testing focus (e.g., concepts and challenges, managing vulnerabilities, security tools and automating security controls) (\textit{Inclusion Criteria 01}). We ended up with 462 webinars in this step.
    \item We then used the video description to select webinars that are centred on rapid deployment or Continuous Software Engineering (CSE) environments (e.g., Continuous practices, DevOps, DevSecOps) (\textit{Inclusion Criteria 02}). Further, we removed webinars that addressed one narrow stage or technological component (e.g., databases) in the overall DevOps or DevSecOps workflows (\textit{Exclusion Criteria 01}). 
    \item  We also performed \textit{quality assessment} of the webinars. While applying the inclusion and exclusion criteria, we browsed through the video to ensure that most of the content of the webinar (i.e., more than half the duration) did not contain marketing material on a commercial product. At the end of this process, 48 webinars were selected for the analysis stage of our study.
\end{itemize}

\subsection{Data analysis}

We selected the \textit{thematic analysis} method \citep{braun2006using} for the data analysis of our study. Thematic analysis enables identifying, analysing, and reporting patterns (e.g., themes) within data \citep{braun2006using, cruzes2011recommended}. This method was selected as we aimed to identify the main themes of challenges, practices and tool features (i.e., our research questions) from a large textual data set.

\begin{figure*} [t]
    \centering
    \includegraphics[width=\columnwidth]{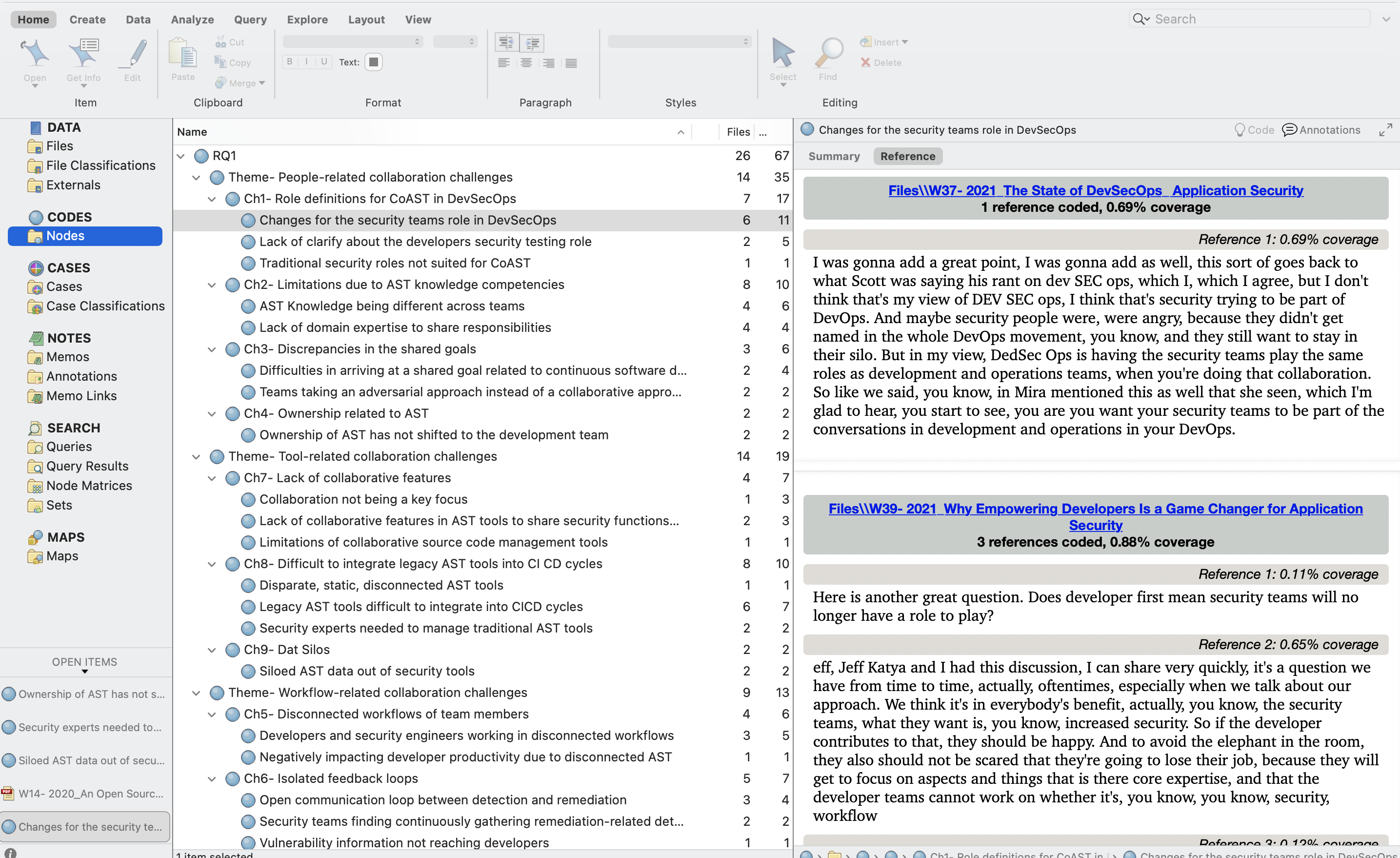}
    \caption{Using Nvivo to create codes in the thematic analysis}
    \label{fig:nvivo}
\end{figure*}

\subsubsection{Familiarising with data}
The first author had been following the selected channels and attending certain webinars (e.g., in DevOpsTV) for over a year. Therefore, the structure or format and typical nature of the webinars were known to the research team.

We started the data extraction process by generating the transcriptions of the selected webinars. As this study was conducted over a period exceeding one year, we experimented with several tools for this purpose. First, we generated transcriptions with a python API, \textit{youtube-transcript-api}\footnote{https://pypi.org/project/youtube-transcript-api/} and then used an AI-based service, \textit{otter.ai}\footnote{https://otter.ai/}, at a later stage. The generated transcriptions were then imported into NVivo, a qualitative data analysis software to conduct \textit{open coding} \citep{glaser1978theoretical, glaser1998doing}. 

\subsubsection{Generating initial codes}
Using NVivo, we created \textit{key points} \citep{hoda2012developing}, which are summarised points of the extracted text from the transcripts. We attempted to create as many key points relevant to our area of interest from the transcript as possible in this step. We then assigned a \textit{code} (i.e., a phrase that further summarizes the key point typically in 2 or 3 words \citep{georgieva2008best}) to the key points. A sample of a code, key point, and the relevant extract from a transcript that resulted in that particular code is shown in (Fig. \ref{fig:nvivo}).

We also captured memos throughout the analysis process using NVivo. These memos were especially helpful for us in capturing informative slides of the webinar presentations (e.g., useful diagrams) and our notes.

\subsubsection{Searching for themes}
We analysed the devised codes and considered how they could be arranged into overarching themes. Accordingly, we drew connections among codes and grouped them into themes using NVivo (Fig. \ref{fig:nvivo}). In this step, we used a multi-layered coding structure in NVivo, to arrange the key points, codes, and themes .

\subsubsection{Reviewing themes, defining and naming themes}
After forming the themes, we considered whether they were coherent, consistent, and distinctive \citep{cruzes2011recommended} (e.g., any overlaps between codes in the themes were rectified). After this review, we finalised the names of the themes in this stage. Finally, we assessed whether the synthesis (i.e., themes and higher-order themes) was adequate to answer our research questions.

\subsubsection{Producing the report}
We detail the findings of our analysis in Section 4. We also present separate tables (Table \ref{table:challenges}, \ref{table:prac}, \ref{table:tools})  for each research question that shows how the themes emerged from the key points of the analysis.

\begin{tcolorbox}[left=1pt, top=1pt, right=1pt, bottom=1pt]
\textbf{Methodological Contribution}: In this study, we managed to capture the challenges faced in the industry, the latest trends and emerging solutions built from the \textit{voices} and \textit{experience} of a number of software practitioners working in a range of roles. This outcome would have been difficult to achieve either through a systematic study of peer-reviewed literature or by setting up practitioner interviews. Therefore, we recommend the systematic process used to analyse the selected data source as a method for capturing the latest areas of interest and practitioner sentiments in emerging technical fields.
\end{tcolorbox}

\begin{figure}[b!]
    \centering
    \includegraphics[scale=.65]{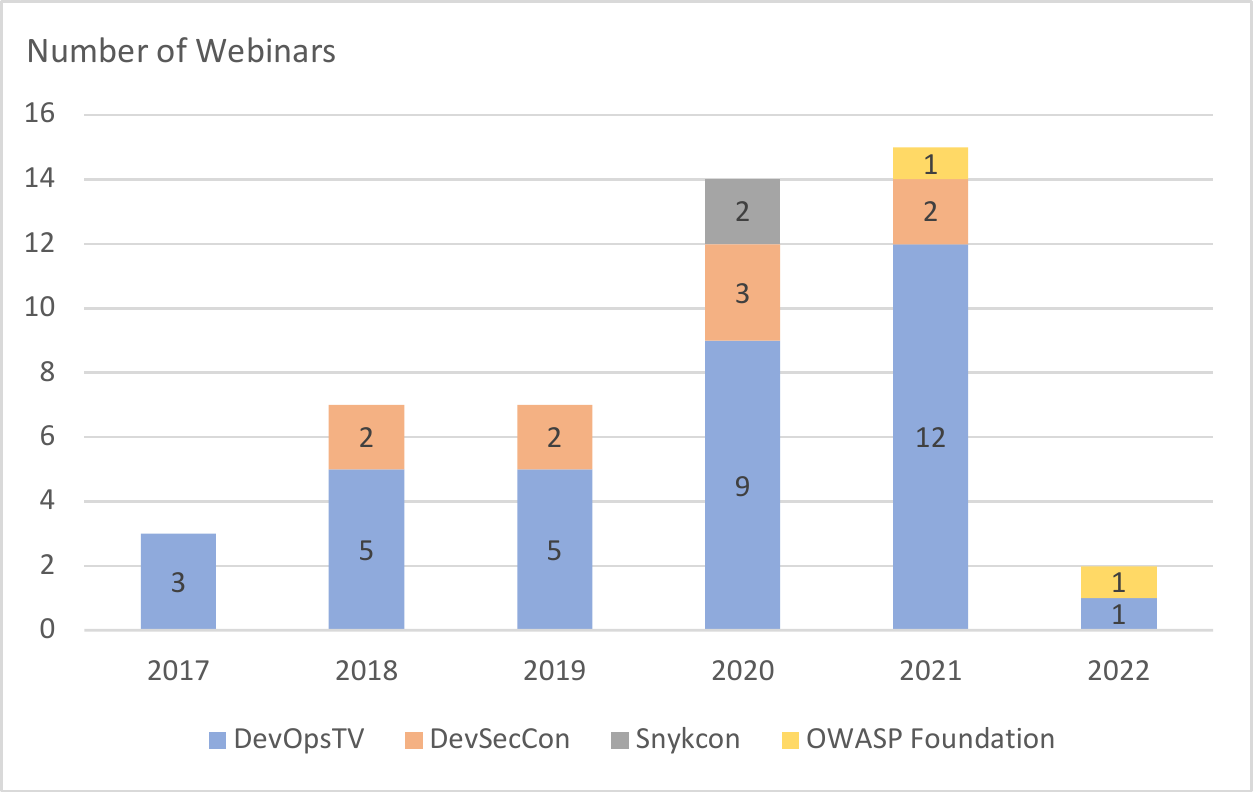}
    \caption{Number of webinars based on the source channel}
    \label{fig:demo}
\end{figure}

\section{Results}

The following sections present the result of our study. We present an overview of the selected webinars and our findings for the research questions.

\subsection{Overview of the selected webinars}
Fig. \ref{fig:demo} shows the number of selected webinars based on the published year (2017-2022) and the source channel. \textit{DevOpsTV} and \textit{DevSecCon} contributed to many webinars for this study (i.e., 35 and 9, respectively). All speakers of the webinars were professionals from the industry. In multiple webinars, more than one person contributed to the presentation. Hence, there was a combined total of 67 speakers. 

Most speakers were employed at leading consulting or AST tool-producing companies (e.g., Synopsys, Veracode, Snyk, WhiteHat Security, Contrast Security, Red Hat, and HCL software). In addition, some webinars were conducted by well-known DevOps industry experts functioning as independent consultants. The speakers identified themselves with titles such as \textit{Security consultants, Solutions or product architects, Research analysts, Transformation consultants, Product engineers or Managers/Directors and DevSecOps Engineers}. There were also many contributions from c-level executives from technology companies (e.g., Chief-technical officers contributed to 9 webinars).  

\begin{table*}[hbt!]
\footnotesize
\caption{Challenges related to performing CoAST in DevSecOps}
  \begin{tabular} {p{.073\textwidth} | p{.22\textwidth}  p{.55\textwidth} |p{.01\textwidth}}
        \toprule 
        \textbf{Theme} & \textbf{Challenges} & \textbf{Key points [Webinars which contributed]}& \#\\ \midrule 

        \textbf{People} 
         & \textbf{Ch1:} Role definitions for CoAST in DevSecOps &     
        \begin{minipage}[t]{\linewidth}
        \begin{itemize}
                    \item Traditional security roles not suited for CoAST [W5]
                    \item Changes for the security teams role in DevSecOps [W17, W18, W27, W28, W37, W39]  
                    \item Lack of clarity about developers' security testing role [W27, W37] 
                    \end{itemize} \end{minipage} &  14
                    \\
                    \cmidrule{2-3} 

        & \textbf{Ch2:} Limitations due to AST knowledge competencies & 
        \begin{minipage}[t]{\linewidth}
        \begin{itemize} 
        
     \item AST knowledge being different across teams  [W9, W15, W20, W36]
     \item Lack of domain expertise to share responsibilities [W5, W8, W24, W38]

    \end{itemize}
    
   \end{minipage} & \\\cmidrule{2-3} 
   
    & \textbf{Ch3:} Discrepancies in shared goals & 
    \begin{minipage}[t]{\linewidth}
    \begin{itemize} 
      
     \item Difficulties in arriving at a shared goal related to continuous software delivery and AST [W15, W24]
     \item Teams taking an adversarial approach instead of a collaborative approach [W15, W36]
    \end{itemize}
    
   \end{minipage} & \\\cmidrule{2-3} 
   
       & \textbf{Ch4:} Ownership related to AST & 
    \begin{minipage}[t]{\linewidth}
    \begin{itemize} 
      
     \item Ownership of AST has not shifted to the development team [W36, W37]
    \end{itemize}
    
   \end{minipage} & \\\midrule

        \textbf{Workf.}
     
         & \textbf{Ch5:} Disconnected workflows of team members &
        \begin{minipage}[t]{\linewidth}
        \begin{itemize}
        \item Developers and security engineers working in disconnected workflows [W15, W23, W28]
        \item Negatively impacting developer productivity due to disconnected AST [W23, W32]
        \end{itemize}
        \end{minipage}
         & 9 \\\cmidrule{2-3}

     & \textbf{Ch6:} Isolated feedback loops & 
                    \begin{minipage}[t]{\linewidth}
        \begin{itemize} 
     \item Open communication loop between detection and remediation [W17, W20, W37]    
     \item Security teams finding continuously gathering remediation-related details difficult [W4, W37]
     \item Vulnerability information not reaching developers [W40]

    \end{itemize}
    
   \end{minipage} & \\ \midrule

\textbf{Tools} & \textbf{Ch7:} Lack of collaborative features &
        \begin{minipage}[t]{\linewidth}
        \begin{itemize}
            \item Collaboration not being a key focus [W21]
            \item Limitations of collaborative source code management tools [W13]
            \item Lack of collaborative features in AST tools to share security functions or data [W16, W34]
        \end{itemize}
        \end{minipage} & 14 \\ \cmidrule{2-3}
        
 & \textbf{Ch8:} Difficult to integrate legacy AST tools into CI/CD cycles &
        \begin{minipage}[t]{\linewidth}
        \begin{itemize}
            \item Disparate, static, disconnected AST tools [W11]
            \item Legacy AST tools difficult to integrate into CI/CD cycles [W7, W8, W18, W22, W32, W36]
            \item Security experts needed to manage traditional AST tools [W4, W8]
        \end{itemize}
        \end{minipage} &  \\ \cmidrule{2-3}     
        
    & \textbf{Ch9:} Data Silos &
        \begin{minipage}[t]{\linewidth}
        \begin{itemize}
            \item Siloed AST data out of security tools [W9, W14]
       \end{itemize}
    
   \end{minipage} & \\\bottomrule
   
\end{tabular}
\label{table:challenges}
\end{table*}

\subsection{RQ1: Challenges related to performing CoAST in DevSecOps}
This section reports the themes of the challenges faced by practitioners in performing CoAST in a DevSecOps environment. The challenges were classified into three main themes: People, Workflow, and Tools (Table \ref{table:challenges}).

\subsubsection{People-related challenges for CoAST in DevSecOps}
A significant source of challenges for CoAST arose from the involved DevSecOps team members (e.g., roles and attitudes). We describe these people-related challenges below. 

\textit{\textbf{Ch1:} Role definitions for CoAST in DevSecOps.}
In the traditional setting, both developers and security engineers have responsibilities that are specific to their function. However, with the move for CoAST, these traditional security roles have become obsolete in DevSecOps [W5]. For example, traditionally, the security team used to define the security policy for the testing tool, enforce the rules, run the tool, and then assess the severity of the outputs [W5]. However, in DevSecOps, with security activities moving to the left of the development cycle, AST is required to be done collaboratively with developers. To achieve this objective, several studies recommend that developers need to play a larger role in these security testing efforts \citep{rajapakse2021empirical, Rajapakse_Zahedi_Babar_Shen_2022a}. Some practitioners are concerned about this recommendation:

\begin{quote}
\textit{Does developer-first (security) mean security teams will no longer have a role to play?} [W39]
\end{quote}

Multiple discussions [W17, W18, W27, W28, W37, W39] raised similar concerns about changes to the role of security teams in DevSecOps. With developers encouraged to do more security testing activities (e.g., SAST), there has yet to be an agreement on the exact role of security engineers in CoAST. Practitioners also seemed to have differing views on which security activities belonged to which party [W27, W37]. 

\begin{quote}
\textit{Oftentimes, I've heard a lot of security people say, Hey, I don't want people to be able to make decisions on risk acceptance or the false positives stuff. And I'm like, Okay, why not? Well, they'll make the wrong decision..!!} [W37]
\end{quote}

Industry-based surveys discussed in [W27] also showed no consensus on who has the primary responsibility for the security testing activities in DevSecOps. When this issue translates into CoAST, team members are confused about their exact role and the specific testing activities they need to perform. 

In addition, the speakers in the analysed discussions offered different advice regarding the security roles of the development and security teams in DevSecOps. The advice ranged from allowing security teams to be involved in day-to-day development activities (e.g., participating in scrum meetings [W28]) to making security a \textit{self-servicing} activity for developers [W18]. Therefore, based on the above details, the need for more clarity about role definitions and responsibilities is a key issue for collaborative work in this domain. 

\textit{\textbf{Ch2:} Limitations due to AST knowledge competencies.} Even though sharing AST responsibilities in DevSecOps is highly recommended in the literature, the lack of domain expertise for team members is a prevalent issue in practice. This is because each team in DevSecOps is typically trained in functionally specific tasks [W9, W15, W20, W36]. For example, it is quite challenging to find engineers with the knowledge of all the relevant security tools and technologies in this context [W20]. Therefore, when a particular AST activity is transferred from one team to another, the team members face difficulties in adequately performing it. 

Both developers and security engineers face limitations related to AST knowledge competencies. As developers are required to run AST tools in their development workflow, they need to know about utilising various security tools. However, practitioners noted that: 

\begin{quote}
\textit{There still exists some pretty huge gaps in application security knowledge and skills amongst developers, and these are being somewhat neglected by organisations.} [W24]
\end{quote}

Interpreting the output of AST tools is one of the main difficulties developers face in this domain. As this task requires expertise related to security vulnerabilities and severity levels, developers without security training find this task challenging [W24]. 

Meanwhile, security engineers are having difficulties adjusting to the new software delivery model of DevSecOps [W9]. Most security team members are trained or familiar with traditional software delivery practices, methods and tools. However, with the introduction of DevOps practices (e.g., continuous practices \citep{shahin2017continuous}) and tools, security personnel are now facing difficulties with the need for knowledge in these concepts to collaborate with development teams adequately. 

The DevSecOps working model requires that all involved team members have a shared goal \citep{mao2020preliminary}. DevSecOps teams, however, have difficulty following this advice [W15, W24]. For example, in this paradigm, developers aim to deliver the software outputs into the production environment as fast as possible. Engaging in AST activities was considered a hindrance to this goal by some developers:

\begin{quote}
\textit{Oftentimes, we see developers seeing traditional security as a blocker. And security, seeing our DevOps teams as trying to push them to the side and move faster than they can ensure security} [W15]
\end{quote}

On the other hand, a thorough security assessment of a software system is the security team's goal. Continuous and rapid delivery of software without comprehensive AST is in direct conflict with that goal. Therefore, these two teams are finding it difficult to arrive at a shared goal in terms of continuous software delivery and AST in this setting. As a result, this has caused teams to take an adversarial approach instead of a collaborative one [W15, W36].

\textit{\textbf{Ch4:} Ownership related to AST.} The ownership of AST traditionally lies with the security team. Nevertheless, development teams must take on some of this ownership as part of the DevSecOps paradigm shift. This shift, however, has not been adequately reflected in practice, according to speakers:

\begin{quote}
\textit{So you cannot say in the industry that the development teams own security, that still hasn't happened. So the ownership still resides with (security teams), like whether it is deciding what tools to include, controls needed or when exceptions are provided. So we are still seeing security teams having more control than the development teams.} [W37]
\end{quote}

Since the development team has minimal control over decision-making processes, this issue discourages them from participating in AST [W36, W37]. For example, resistance to AST might occur when developers do not have a say on the specific tools they prefer to use within their work environment [W36]. Therefore, more effort needs to be directed at how developer teams can be empowered to have more ownership related to AST while engaging in coding.

\subsubsection{Challenges related to workflows in CoAST}
In a software development environment, each team has its own \textit{workflow} to produce or add value to the software outputs. This section reports the challenges related to these workflows in performing CoAST.

\textit{\textbf{Ch5:} Disconnected workflows of team members.} For CoAST to be successful, the workflows of DevSecOps team members need to be adequately integrated \citep{rajapakse2021empirical}. However, the practitioners' survey results discussed in [W23] showed a disconnect between developer and security teams' workflows.

\begin{quote}
\textit{96\% of developers report that the disconnect between their own workflows and that of the security team is actually hurting their productivity [..] when you see a signal this strong, you really know that you're on to something.} [W23]
\end{quote}

The speakers highlighted various examples of how the above-noted disconnect occurs in practice. A key contributing factor for this disconnect arises from traditional AST and modern continuous practices-based processes occurring in separate workflows [W23]. AST is typically performed by security teams during the build or test stages [W23]. These tests are usually performed on multiple feature commits for a fixed period (e.g., a day). If a security team uncovers any vulnerabilities from the AST, they direct them to the development team based on the severity level. However, speakers point out many drawbacks of this practice (i.e., teams working separately on their own workflows). For example, considering that the security team provides the vulnerability report sometime after the commit was done, the developer has to choose between continuing to work on the next item (e.g., a new feature) or fixing the security issues [W32]. This impacts the developers' productivity [W23, W32]. 

\textit{\textbf{Ch6:} Isolated feedback-loops.} Continuous communication between development and security teams is a key recommendation for DevSecOps \citep{sanchez2020security, mao2020preliminary}. In practice, however, such an unrestricted flow of information was described as a challenge [W4, W17, W20, W37, W40]. For example, speakers discussed difficulties related to relevant security information not reaching both the development and security teams:

\begin{quote}
\textit{[..] security team will tell you to fix these 15 vulnerabilities. And then the developer would go, well, maybe this one is a false positive, [..] this one I cannot fix because I cannot upgrade the version. And then you have this feedback loop, this back and forth [..] (but) somehow the information is not flowing back to the security teams, and I think this is one of the challenges, closing this loop} [W37]
\end{quote}

In AST, the detection of vulnerabilities is comparatively straightforward; however, the remediation part is challenging as it would require collaboration between several parties:

\begin{quote}
\textit{If you're using a dedicated tool [..], detection of vulnerabilities is pretty easy. But the hard thing is to close the loop and actually remediate those vulnerabilities.} [W17]
\end{quote}

As stated in [W17], the security teams find it challenging to continuously gather important remediation-related details, such as whether a developer actually fixed the issue, the nature of the remedy and the vulnerabilities remaining to be addressed. Further, if the expectation is for developers to engage in AST, the relevant information needs to be continuously provided to them. However, the vulnerability-related data not reaching developers is a challenge discussed concerning these feedback loops [W40]. The speakers highlighted that email and ticket-based workflows in the silo-centric work culture contribute to this situation [W17, W20, W37].

\subsubsection{Tool-related challenges for CoAST in DevSecOps}
DevSecOps is an environment where tools are heavily used due to the need for automation for rapid deployments. Collaborative work in this area, however, is challenged by several tool-related challenges that we report below.

\textit{\textbf{Ch7:} Lack of collaborative features.} For effective collaboration, tool support is considered a useful \textit{enabler} \citep{Rajapakse_Zahedi_Babar_Shen_2022a}. For example, there are many source code management (SCM) tools that allow effective and efficient collaboration, such as Git \citep{git_2022}. However, such collaborative SCM tools normally do not have the required security features [W13]. As a result, developers are compelled to look for separate security tools or frameworks for these tasks [W13]. But the problem is that many such security tools lack features that facilitate inter-team collaboration [W16, W21]. This issue particularly affects developers who are required to do AST while seeking regular inputs from a security team member.

In DevSecOps, there is a requirement to share code scanning results and related information efficiently across teams [W34]. In addition, sharing the particulars of a code problem (e.g., code snippets) is required in certain situations. Unfortunately, most available AST tools are yet to focus on this requirement for collaborative work, even though it is vital for DevSecOps [W13]. 

Another example discussed in [W16] is the lack of proper support for developers to manage the frequently occurring false positives of AST tools: 

\begin{quote}
\textit{Overall, security tools haven't usually very much thought about the developer experience, right? So a lot of tools claim to be DevSecOps friendly, but you need to log into a portal for pointing or clicking to manage your false positives, right? And that's the lack of integrated thinking into letting developers manage false positives.} [W16]
\end{quote}

Therefore for CoAST, the lack of adequate tool support is a significant concern, particularly due to the tool-centric nature of DevSecOps.

\textit{\textbf{Ch8:} Difficult to integrate legacy AST tools into CI/CD cycles.} 
In a DevSecOps environment, developers work in continuous integration/continuous deployment (CI/CD) cycles. Thus, security tools are required to integrate into these cycles (or \textit{pipelines}) for effective collaboration with developers. However, many legacy tool drawbacks were discussed in the webinars that made tool integration difficult:

\begin{quote}
\textit{If you look at the state of the practice that people have employed to date for application security, you get what we call this legacy tool quagmire. You have a set of disparate, static, disconnected tools that, taken together, are largely inaccurate} [W11]
\end{quote}

As a result, developers are finding it challenging to integrate many types of AST tools into their development pipeline [W7, W8, W18, W22, W36]. In [W18], a tool producer pointed out a case of one of their customers taking as long as two years to integrate a security tool into their pipeline. These factors ultimately make developers unwilling to use these tools to perform AST.

The speakers also noted that the legacy AST tools are not developed for modern CI/CD practices [W32]. Most such tools are catered towards a \textit{one point of time} scan (e.g., build phase) [W7] and would not be capable of providing continuous scanning functionality for developers [W32]. 

AST tools have also historically been targeted at security professionals. Therefore, they require considerable security expertise to use [W4]. As a result, even if these tools are introduced in CI/CD workflows, a developer would find it difficult to get the most use out of them [W8]. 

\textit{\textbf{Ch9:} Data silos.} In a typical DevSecOps pipeline, software developers, security engineers and operators use many tools. This usually includes different types of security testing tools during various stages in the pipeline. The problem arises when the number of tools becomes high, and the product type differs substantially. In this scenario, getting the maximum use out of the tools becomes challenging. Further, collaboration across teams gets quite difficult when each team uses its own tool sets, which do not integrate in any manner. A significant concern that results from this situation is the siloed security data from disparate security tools [W9, W14]:

 \begin{quote}
\textit{[..] if you implement software to perform static analysis, that data cannot be used by another software that does static analysis or another software that does, say dependency or container scanning, the data will become siloed into the product or platform} [W14]
 \end{quote}

Such data silos are problematic, especially when developers do the testing and require feedback from the security team to assess some part of the output. The speakers pointed out that AST tools being developed by separate companies using different roadmaps without aiming for a collaborative testing approach causes this issue [W14].

\begin{tcolorbox}[left=1pt, top=1pt, right=1pt, bottom=1pt]
\textbf{RQ1 Summary:} Although DevSecOps is centred on collaboration, limited attention is paid to people-centric issues, such as role definitions, establishing shared goals, and transferring ownership of AST, making CoAST difficult in practice. Additionally, such challenges are exacerbated by the absence of collaborative features and other limitations in AST tools. Ultimately, practitioners face a trade-off between increased deployment speeds, versus AST in this environment.
\end{tcolorbox}

\subsection{RQ2: Recommended best practices to perform CoAST in DevSecOps}
This section reports the best practices recommended by practitioners to perform CoAST in a DevSecOps environment. The practices were thematically classified into five main themes described below (Table \ref{table:prac}).

\begin{table*}[htbp]
\footnotesize
\caption{Best practices recommended to perform CoAST in DevSecOps}
  \begin{tabular} {p{.12\textwidth} | p{.18\textwidth}  p{.57\textwidth} |p{.01\textwidth}}
        \toprule 
        \textbf{Theme} & \textbf{Practices} & \textbf{Key points [Webinars which contributed to the point]}& \#\\ \midrule 

        \textbf{Shift in practices} 
         & \textbf{P1:} Shift-left application security testing &     
        \begin{minipage}[t]{\linewidth}
        \begin{itemize}
                    \item Enabling better collaboration using shift-left [W5, W9, W17, W23]
                    \item Enables shared tooling [W5]
                    \item Shift security from a reactive to a proactive posture [W5]
                    \end{itemize} \end{minipage} &  
                    6 \\
                    \cmidrule{2-3} 

  & \textbf{P2:} Shift-right security in production &
        \begin{minipage}[t]{\linewidth}
        \begin{itemize}
        \item Scanning in production environments [W7, W9]
        \item Deploy and monitor security controls in production [W27]
        \end{itemize}
        \end{minipage}
         &  \\\midrule

        \textbf{Team structures}
     
         & \textbf{P3:} Forming hybrid or cross-functional teams &
        \begin{minipage}[t]{\linewidth}
        \begin{itemize}
        \item Consists of cross-functional team members [W29]
        \item Functional members have an overall view [W31]
        \item Cross-disciplinary teams having shared ownership [W31]
        \end{itemize}
        \end{minipage}
         & 14 \\\cmidrule{2-3} 
         
     & \textbf{P4:} Changing traditional software development roles & 
                    \begin{minipage}[t]{\linewidth}
        \begin{itemize} 
     \item Developers expanded role on security testing [W1, W4, W15, W23, W24, W25]
     \item Governance role for the security team [W4, W15, W28]
     \item Security engineers familiarising with DevOps practices [W9]

    \end{itemize}
    
   \end{minipage} & \\ \cmidrule{2-3} 
   
        & \textbf{P5:} Assigning security and DevOps champions &
        \begin{minipage}[t]{\linewidth}
        \begin{itemize}
        \item Training a developer to focus more on security [W4, W30, W35, W36, W37]
        \item Embed a DevOps champion within the security team [W36]
        \end{itemize}
        \end{minipage}
         &  \\\midrule

\textbf{Access Management} & \textbf{P6:} Collaborating using pull requests &
        \begin{minipage}[t]{\linewidth}
        \begin{itemize}
            \item  Managing pull-requests to access code repositories [W2, W13, W15, W23, W32, W39]
            \item Collaborative ways for AST using pull-requests [W2, W13, W23]
            \item Enforcing organisational security standards or thresholds [W15, W23]
        \end{itemize}
        \end{minipage} & 8 \\ \cmidrule{2-3}
        
 & \textbf{P7:} Role-based access policies &   
        \begin{minipage}[t]{\linewidth}
        \begin{itemize}
            \item Role-based access control for the code and pipeline [W15, W16]
            \item Centrally managed policies [W20]
        \end{itemize}
        \end{minipage} &  \\ \midrule

   \textbf{Data-centric collaboration} & \textbf{P8:} Centralised security data repositories &
        \begin{minipage}[t]{\linewidth}
        \begin{itemize}
            \item Centralised repositories for AppSec tool data [W19, W36]
            \item Holistic view of the security status [W36, W38]
            \item Making vulnerabilities data useful and operational across teams [W36]
            \item Normalise, correlate and prioritise centralised data [W36]      
        \end{itemize}
        \end{minipage} & 7 \\ \cmidrule{2-3}

    & \textbf{P9:} Vulnerability data sharing practices &
        \begin{minipage}[t]{\linewidth}
        \begin{itemize}
            \item Coordination of data sharing practices [W33]
            \item Combine and curate industry standard vulnerability data [W33]
            \item Bug bounty programs to collaborate with external parties [W3, W23, W40]
       \end{itemize}
    
   \end{minipage} & \\\midrule
   
      \textbf{Comm. practices} & \textbf{P10:} Continuous feedback loops &
        \begin{minipage}[t]{\linewidth}
        \begin{itemize}
            \item Instill a culture of learning to seek continuous input [W4]
            \item Feedback to be treated as educational [W42]
            \item Shorter feedback loops [W14, W42]
            \item Enables better visibility [W4]
        \end{itemize}
        \end{minipage} & 6 \\ \cmidrule{2-3}
        
 & \textbf{P11:} ChatOps &
        \begin{minipage}[t]{\linewidth}
        \begin{itemize}
            \item A workflow that integrates people interactions and tools within communication platforms [W47]
            \item Instant feedback about security issues [W11]
            \item Improved transparency and inter-team collaborations  [W6, W11, W47]
        \end{itemize}
        \end{minipage} &  \\ \bottomrule  
        
\end{tabular}
\label{table:prac}
\end{table*}

\subsubsection{Shifts in the practices}
In the traditional software development setting, each team had functionally specific practices. However, to enable collaborative work in DevSecOps, certain functional practices were recommended to be shifted along the development process. As a result, members of different functional teams were encouraged to share work through a collaborative workflow which includes the following best practices.

\textit{\textbf{P1:} Shift-left application security testing.} Traditionally, AST is considered an activity to the \textit{right side} of the development process, done by the security team. However, this resulted in functional silos in terms of development practices and AST (i.e., AST done asynchronously at a later stage, separate from the coding). To overcome this situation and to enable better collaboration, shifting AST to the left (i.e., \textit{shift-left security}) of the cycle was a frequently noted recommendation in the webinars [W5, W9, W17, W23]. 

However, speakers noted that the practical implementation of this practice would depend on the team size [W9]. For example, in a larger organisation, shift-left would entail developers frequently collaborating with the security teams (e.g., corporate security, IT security, and cyber security teams) earlier in the process. This would also require security teams to engage in DevOps practices performed by developers (e.g., continuous practices). Both these shifts would mean a high amount of collaboration between these teams:

\begin{quote}
\textit{So in a way, we're kind of blending disciplines in Shift-left. And I think the most important thing that I take away from this evolution is it's about collaboration, and it's about people.} [W9]
\end{quote}

As a result of shift-left security practices, there were several significant advantages for collaboration in DevSecOps environments. Firstly, team members would be required to share the tooling they use in their functions. For example, developers would get the opportunity to use security tools while engaging in their development practices. In this case, the goal is to present security information to developers in their own environment [W5]. Secondly, these practices would create a closed loop of regular collaboration between these teams [W17]. This would result in greater awareness of each team's practices and more empowerment for cross-collaborative work. Ultimately this best practice shifts security from a reactive to a proactive security posture in a rapid deployment environment [W5]. 

\textit{\textbf{P2:} Shift-right security in production.} In contrast to shift-left (i.e., P1), shift-right security involves shifting code scanning practices and security controls to the right of a typical development process. In the webinars, tool producers noted that customer interest is gaining for these practices for rapid deployment environments [W7, W9]. In DevSecOps, the codebase in production is regularly changing. Due to rapid deployment needs, certain code segments could also be left untested during rapid deployments. This could result in vulnerable code in production. Therefore, practitioners recommended shifting certain code scanning (e.g., DAST ) [W6, W7] and runtime protection controls [W27] to the production environment (i.e., shifting right).

\begin{quote}
\textit{DAST scanning is an integral part of a well-rounded AppSec program. When you do dynamic scanning in runtime, the vulnerabilities that are found are actionable. Those are the ones that the hackers will take advantage of. So DAST scanning can be done both in pre-production and production environments.} [W7]
\end{quote} 

Similar to shift-left resulting in developer and security team collaboration (e.g., P1),  shift-right offers an opportunity for development and security teams to collaborate with the operations team that manages the software in production. Ultimately when both above-mentioned shifts in practices (i.e., P1 and P2) are combined, it would ensure complete application security coverage for the software [W7].

\subsubsection{Team structures}
This section presents the synthesis of best practices to modify existing traditional team structures or functional roles to aid CoAST in DevSecOps. 

\textit{\textbf{P3:} Forming hybrid or cross-functional teams}. 
We previously discussed several people-related challenges (e.g., Ch1-C4) for CoAST stemming from team members working in functionally specific units. To address these problems, speakers proposed forming \textit{hybrid teams} for DevSecOps environments [W29, W31]. We extracted several characteristics of such teams in the webinar discussions.

Firstly, hybrid teams should consist of cross-functional team members involved in DevSecOps: 

\begin{quote}
\textit{[] you have all the different pieces (team members) you need to get software out the door without having to go to another department. So in terms of the security aspect, you want to embed security (team members) into those teams. So you're not reaching out every five minutes to the security department on checks a, b, and c.} [W29]
\end{quote}

Such cross-disciplinary teams would streamline the application development pipeline (e.g., tools and technologies used) and create an efficient workflow agreed upon by all members [W29]. Speakers also advised that hybrid team members should have an overall view of the process and have shared goals (as opposed to functional objectives or milestones [W31]). This advice is also beneficial in reducing conflicts that may arise in collaborative work. For example, in [W31], it was noted that cross-disciplinary teams would be a solution for the shared ownership problem (i.e., developer teams unwilling to take security ownership and lack of clarity about who exactly owns security in DevSecOps). 
 
\textit{\textbf{P4:} Changing traditional software development roles.} Developers engaging in AST in collaboration with the security team was one of the frequently repeated best practices for DevSecOps [W1, W4, W15, W23, W24, W25]. The main challenge regarding this proposal was developers lacking the required security expertise to engage in AST actively (e.g., Ch2). Nevertheless, speakers noted that this recommendation does not require developers to turn into security experts: 

\begin{quote}
\textit{We don't think that developers need to be able to become penetration testers. But by understanding enough of the basics of how an application may be attacked, you can properly defend against it, you can properly leverage the results of automated and other types of security testing to make yourself more secure. That's an important piece of the puzzle.} [W1]
\end{quote}

What is important, however, is to close the security skills gap between the development and security teams via frequent collaboration and adequate workflows or practices [W24]. For example, in this role change, developers need to use the automated testing tools previously used by the security team at specific checkpoints of the development process [W40]. The assessment of the tool output (e.g., the severity level of vulnerabilities) can be done in collaboration with the security team. 

On the other hand, the security team's role has transformed into a governance role, reducing operational security workload [W4, W15, W28]. This includes tasks such as setting up the security policies, guiding the developers and having visibility (i.e., monitoring) of the development and operations processes [W15, W28]. In this new role, the security team needs to have the ability to answer questions such as: 

\begin{quote}
\textit{How good are we progressing (e.g., security operations)? How fast did it take the development teams to remediate vulnerabilities? Are we getting better or worse at handling the security aspect?} [W15] 
\end{quote}

Finally, to successfully operate in this environment, speakers also mentioned the necessity for security engineers to be knowledgeable of development-related DevOps practices [W9]. 

\textit{\textbf{P5:} Assigning security and DevOps champions}.
A \textit{security champion} is a developer who pays special attention to the security aspects of the development process and software outputs [W4, W30, W35, W36, W37]. Such security champions have an active role in communication, knowledge sharing, and collaboration among DevSecOps team members [W30]. While a security champion can aid other developers more frequently (as they are part of the same development team), they can also act as an intermediary with the security team. Speakers also emphasised that such developers need not be experts in all aspects of security [W37]. Nevertheless, certain speakers emphasised that a security champion should know the limits of their knowledge, know when to escalate and seek guidance from the security team [W4]. 

\textit{DevOps champions} is another role suitable for collaboration in this context discussed in the webinars [W36].

\begin{quote}
\textit{I believe that for large organisations, it could make just as much sense to embed a DevOps champion within the security team. (This is) someone who understands the day-to-day process and all the different types of tools that the DevOps teams are using.} [W36]
\end{quote}

Like security champions, this role intends to bridge the communications gap between the development or DevOps teams and the security team. 

\subsubsection{Access Management}
DevSecOps recommends wider access to the code and tool pipeline to enhance collaboration. The following best practices address the secure implementation of this recommendation.

\textit{\textbf{P6:} Collaborating using pull requests}. 
Broad access to code repositories is a prerequisite for CoAST work in this context. However, clearly defined rules and practices should be in place to facilitate this collaboration [W13]. In modern software development environments, distributed version control (DVC) systems (e.g., Git \citep{git_2022}) based coding practices are widespread. \textit{Pull-requests} (i.e., a merge request to the base branch for which the requester does not have write permissions) were proposed by speakers as a method of collaborating and controlling access in these environments. The proposed workflow in several webinars [W23, W32, W39] is as follows. 

When the code is ready, developers will create a pull request to merge one or more commits into the base branch. Automated security testing and other unit tests can be performed at this point, and the security team can review results before the pull request is approved and the code is merged [W15, W23, W32]. If there are security vulnerabilities present at this point, the developer can get the assistance of the security team to discuss the implications and rectify them immediately. 

Several benefits of these collaborative AST practices were discussed: 

\begin{quote}
\textit{It's easy for developers as all happens within their environment, they get this feedback immediately, they don't have to wait around. And they get the feedback when it's most relevant.} [W23].
\end{quote}

Speakers emphasised that this is a best practice for CoAST in this setting as it would lead to fewer conflicts [W23]. Because of the advantages mentioned above, the developer is in a better position to address vulnerabilities if the security team denies their pull request.

In addition, organisations also get the opportunity to enforce security standards, thresholds or \textit{phase gates} before changes are made in code repositories [W15, W23]. Here, speakers noted that developers should not \textit{just scan code for the sake of scanning code} but continually and deliberately ensure that the proper tests are done in collaboration with the security team [W15]. This best practice also ensures that the code must pass established phase gates for it to move further down the pipeline [W15]. 

\textit{\textbf{P7:} Role based access}.
There is an opposing view for collaborative work in this context due to fears that broader access to the code and technical pipeline may result in intentional (e.g., insider threats \citep{Rajapakse_Zahedi_Babar_Shen_2022a}) or unintentional damages (e.g., data loss \citep{10.1145/3558001}). To address this issue, speakers recommended forming clearly defined role-based access policies as a best practice [W15, W16].

\begin{quote}
\textit{It's important to establish role-based access controls and permissions as well as checks on those permissions to ensure that the right people are making the right changes [..] You need to ensure that team members don't have access to change portions of the overall technical stack and compromise security.} [W15]
\end{quote}

In role-based access, the advice is to create specific roles for the team members based on their collaborative needs. The specific permission-related details of these roles should be included in centrally managed security policies [W20].

\subsubsection{Data-centric collaboration}
Ch9 in our study focused on the data-centric challenges encountered in CoAST. The following practices were recommended to address this specific challenge.

\textit{\textbf{P8:} Maintaining centralised security data repositories.}
For developers, AST presents a significant challenge of making sense of data generated by multiple tools that different DevSecOps team members use. One strategy proposed to mitigate this issue was storing and maintaining diverse data sets from such tools in a centralised repository [W19, W36]. Tools producers further noted that these centralised data need to be normalised and correlated (due to the diverse data sources or tools): 

\begin{quote}
\textit{We're going to ingest all of that data into our centralised repository. From there, we're going (to perform) data refinement, normalise the data into a common risk framework, compress, remove noise or duplicates and correlate data across tools. What this is going to do is allow developers to focus on the most meaningful things} [W38]
\end{quote}

Such a practice is beneficial in providing developers with prioritised security data to focus their efforts.

From the collaborative perspective, centralised data repositories enable a holistic view of the data across the teams involved in DevSecOps [W36, W38]. For example, when developers store the AST results obtained while running the tools in their own environments in a centralised location, the security team would have better visibility of the data [W36]. Therefore, a centralised data strategy would ultimately make vulnerability information useful and operational for the involved teams [W36]. 

\textit{\textbf{P9:} Vulnerability data sharing practices}. 
Vulnerability data in the industry is rapidly growing, with many security vulnerabilities being discovered every day. Therefore, government bodies and companies have taken steps toward creating publicly accessible databases to manage and disseminate this data (e.g., The National Vulnerability Database (NVD) [W33]). In addition to these databases, there are other information sources that need to be regularly checked for vulnerability information (e.g., \textit{we look at bug or issue trackers, GitHub, or security advisories coming out of Microsoft} [W33]). Speakers discussed that information from such diverse sources needs to be combined and curated regularly [W33]. Typically, these tasks are done by the security team members. Therefore, they need to coordinate data-sharing methods (e.g., data pipelines) or practices to share the data with the development team on time [W33]. 

Another source of vulnerability information is \textit{bug bounty programs} [W3, W23, W40]. Such programs were highlighted as a method to collaborate with external parties to perform security assessments and provide useful vulnerability data.

\subsubsection{Communications practices}
This section presents the communications-related best practices recommended for CoAST in DevSecOps.

\textit{\textbf{P10:} Continuous Feedback loops}.
As the software delivery in DevSecOps occurs in continuous cycles, continuous feedback loops with all involved teams are a best practice in this environment. However, such feedback loops being restricted was highlighted in the webinars as a challenge for practitioners (e.g., Ch6). To overcome this problem, speakers emphasised the need to instil a \textit{culture of learning} in the involved teams so that they are inclined towards seeking continuous input [W4]. 

\begin{quote}
\textit{So the feedback loop, the amplifying feedback is incredibly close to when you wrote it (code). And now I can use this (security test) as almost like an educational scan. Oh, I just wrote something that's broken. Let me go and fix it.} [W42]
\end{quote}

The aim here is to treat continuous feedback as educational as opposed to contentious. Such a practice would encourage both team members to offer and receive feedback. 

Several advantages of continuous feedback loops were discussed in the webinars [W4, W16, W42]. For example, continuous feedback loops aid in quality checks based on all the continuous changes made in the paradigm. Further, regular communication cycles increase operational visibility and collaboration across teams [W4]. 
 
\textit{\textbf{P11:} ChatOps}. \textit{ChatOps} is an emerging collaboration model for \textit{conversation-driven collaboration}, ideal for complex environments such as DevSecOps. This model proposes a workflow that integrates people, their communications, documents, tools and processes into a collaborative environment [W47]. We captured several mentions of ChatOps in the webinars [W6, W11, W47]. 

One example of a ChatOps implementation is integrating tools used in a DevSecOps environment within a communications platform (e.g., Slack) [W47]. The overall aim of such an implementation is to create an environment where DevSecOps team members receive instant information, such as security feedback  [W11, W47]:

\begin{quote}
\textit{Developers are armed with this always-on 24 by 7 vulnerability discovery engine that empowers them to check-in cleaner code at the very beginning of the application development lifecycle.} [W11]
\end{quote}

As a result of adopting the ChatOps practices, speakers noted several benefits such as improved knowledge management, transparency and increased inter-team collaborations [W6, W11, W47].

\begin{tcolorbox}[left=1pt, top=1pt, right=1pt, bottom=1pt]
\textbf{RQ2 Summary:} The webinars proposed a range of best practices for CoAST, addressing major pain points in the context, such as the silo-based work culture. Many practices targeted enabling DevSecOps team members the required security when it is most useful (i.e., \textit{access to the right help at the right time}, e.g., forming hybrid teams, ChatOps, continuous feedback loops).
\end{tcolorbox}

\subsection{RQ3: Tool types and capabilities that support CoAST in DevSecOps}

In this section, we present the extracted tool types and capabilities that support CoAST in a DevSecOps environment (Table \ref{table:tools}). 

\subsubsection{Integration}

\textit{\textbf{T1:} Integration with developers platforms.} Team members use many different tools in a DevSecOps environment. Effective collaboration in this context requires robust integration between these tools used by different teams. For example, selecting AST tools that can integrate with developer platforms was a frequently noted recommendation presented by speakers [W14, W15, W23, W27]. Here, the aim is to enable developers to rectify security issues in a platform or environment familiar to them [W23]. The security team's role would be to assist the developer in this process.

However, a current challenge for CoAST is to shift legacy tools, which target security professionals [W23], to the developers working environment. This situation has given rise to a new generation of \textit{developer-centric} AST tools that integrate into the developers' workflow, as opposed to targeting security professionals [W15, W23, W27].  

A key feature captured on developer-centric tools is the ability to integrate into industry-standard Integrated Development Environments easily (IDEs). Such AST tool integration with the IDE would provide instant security feedback enabling security vulnerabilities to be rectified while the developer is writing the code [W44]. 

\begin{quote}
\textit{So we chose to actually start with the IDE integration specifically because we believe that we provide the greatest value to the developer (on the IDE). And we empower them (so that) they can do security testing themselves.} [W43]
\end{quote}

Similar to IDEs, AST tools should have capabilities to integrate with other highly used development-related tools such as source code management systems or build engines. Notably, AST tools need to integrate with development environments because if developers are required to move away from their workflow to perform security testing, there could be resistance. This is due to developers' preference to work with tools and processes familiar to them [W44]. Accordingly, integrating with existing development tools and platforms is a key capability to consider when selecting an AST tool.

\begin{table*}[htbp]
\footnotesize
\caption{Tool types and capabilities that support CoAST in DevSecOps}
  \begin{tabular} {p{.18\textwidth} | p{.2\textwidth}  p{.55\textwidth} |p{.01\textwidth}}
        \toprule 
        \textbf{Theme} & \textbf{Tool type or capability} & \textbf{Key points [Webinars which contributed to the point]}& \#\\ \midrule 

        \textbf{Integration} 
         & \textbf{T1:} Integration with developers platforms &     
        \begin{minipage}[t]{\linewidth}
        \begin{itemize}
                    \item Most legacy AST tools target security professionals [W23]
                    \item Developer-centric AST tools that integrate into the developers' workflow as a key capability [W14, W15, W23, W27].  
                    \item  Ability to easily integrate into industry-standard Integrated Development Environments (IDEs) [W15, W43]
                    \item Instant security feedback [W44]
                    \end{itemize} \end{minipage} & 15 
                    \\
                    \cmidrule{2-3} 

    & \textbf{T2:} Integration with other security tools or sources & 
    \begin{minipage}[t]{\linewidth}
    \begin{itemize} 
      
     \item The need for diverse AST tool integration [W11, W14, W30]
    \item APIs to integrate with the tool pipeline [W4, W12, W36, W46]
    \item APIs to pull vulnerability information from diverse sources [W8]
    \item Tool orchestration methods that enable integration [W29, W36, W45]
    \end{itemize}
    
   \end{minipage} & \\\midrule 

  \textbf{Visualisation} & \textbf{T3:} Visualisation of AST information &
        \begin{minipage}[t]{\linewidth}
        \begin{itemize}
        \item Visualisations to communicate vulnerability information across teams [W14, W48]
        \item Compatible tool reports [W14]
        \end{itemize}
        \end{minipage}
         & 2 \\\midrule
         
  \textbf{Configuration} & \textbf{T4:} Flexibility in configurations and rules &
        \begin{minipage}[t]{\linewidth}
        \begin{itemize}
	\item The need for accuracy in the output [W23]
        \item Offers flexibility in configuring the tool or rule sets collaboratively [W4, W14, W19]
        \end{itemize}
        \end{minipage}
         & 4 \\\midrule 
         
      \textbf{Continuous communication} & \textbf{T5:} Communication platforms  &
        \begin{minipage}[t]{\linewidth}
        \begin{itemize}
            \item Transferring vulnerability information to developers [W8]
            \item Facilitating feedback loops [W46]
        \end{itemize}
        \end{minipage} & 5 \\ \cmidrule{2-3}
        
        & \textbf{T6:} Team collaboration software &
        \begin{minipage}[t]{\linewidth}
        \begin{itemize}
            \item Storing and sharing AST-related process models [W19]
            \item Sharing engagement models across teams [W20]
        \end{itemize}
        \end{minipage} &  \\ \cmidrule{2-3}
        
          & \textbf{T7:} Automated Bots &
        \begin{minipage}[t]{\linewidth}
        \begin{itemize}
            \item Automated bots to provide continuous feedback [W47]
       \end{itemize} \end{minipage} & \\ \midrule      
        
       \textbf{Hybrid testing} & \textbf{T8:} Interactive Application Security Testing (IAST) &
        \begin{minipage}[t]{\linewidth}
        \begin{itemize}
        \item  Uses software instrumentation to provide continuous security [W6, W8, W11]
        \item Facilitates a better collaborative AST workflow [W8] 
        \end{itemize}
        \end{minipage}
         & 3 \\\midrule         
        
 \textbf{Tracking and Correlation} & \textbf{T9:} Issue or bug tracking systems &
        \begin{minipage}[t]{\linewidth}
        \begin{itemize}
            \item For coordination between developers and security teams [W4, W44]
             \item For managing the vulnerability remediation in CoAST [W8, W11, W33, W34]
        \end{itemize}
        \end{minipage} & 8 \\ \cmidrule{2-3}     
        
    & \textbf{T10:} Security Information and Event Management (SIEM) &
        \begin{minipage}[t]{\linewidth}
        \begin{itemize}
            \item Manages collaboration of disparate tools of different teams [W48]
            \item Provides actionable outputs [W36, W48]
       \end{itemize}
    
   \end{minipage} & \\\bottomrule
   
\end{tabular}
\label{table:tools}
\end{table*}

\textit{\textbf{T2:} Integration with other security tools or sources.} A DevSecOps environment typically includes different types of AST tools in various stages in the pipeline. However, when the number of tools becomes high and the nature of the products differs, getting the maximum use out of these tools becomes challenging [W30]. Further, collaboration across teams becomes quite difficult when each team uses its own AST tools, which do not integrate in any way. 

Speakers, therefore, emphasised the need for AST tools that can be easily integrated with other security tools [W11, W14]. Different technologies or services could perform this integration. However, the AST tool must have the required flexibility or features for this integration. A key feature in this regard is the availability of Application Programming Interfaces (APIs). For example, an API use discussed in the webinars was interfacing AST tools with the DevOps or DevSecOps tool pipeline [W4, W12, W36, W46]. Here, practitioners highlighted the importance of using APIs of AST tools used by different teams to integrate them into a single robust pipeline [W4, W46]. Such an integrated tool pipeline across teams is beneficial in facilitating inter-team collaboration.

Some tool producers also noted that APIs are an efficient way to collaborate with external security data providers. For example, several prominent vulnerability-related security information sources are available. These sources are updated frequently, and traditionally this information is handled by the security team. However, if these sources can be directly interfaced with AST tools used by developers, critical security information could reach them without delay [W8]. 

Another method to achieve diverse tool integration is the use of orchestration services [W29, W36, W45]. For example, in [W45], the capabilities of \textit{intelligent orchestration} was discussed:

\begin{quote}
\textit{So intelligent orchestration is a way to take the events that occur within the development pipeline and feed them into an intelligent system that looks at the context in which that event occurred. And based on policy, decide what kind of test should be run? Should I do static analysis on code? Should I do dynamic analysis on an API or a binary?} [W45]
\end{quote}

Having such orchestration methods to manage different tools would direct developers and security teams to a common platform, enabling better CoAST [W45]. 

In the traditional setting, developers found it problematic when the security team came to them at the last moment requesting security fixes to the code. However, the integration-based tool capabilities mentioned in this section would give the developers more control as they can use preferred AST tools in their own work environment and be more aware of important security updates as and when they are available. Such measures would ultimately lead to a better collaborative relationship with the security team instead of an adversarial one.

\subsubsection{Visualisation}

\textit{\textbf{T3:} Visualisation of AST information} 
Speakers discussed the need for effective visualisation features in AST tools for better collaboration. Due to the large amount of data generated by AST tools, visualisation features (e.g., reports and dashboards) have a crucial role in facilitating collaborative work in DevSecOps [W14, W48]. Several AST tool producers who spoke in the webinars presented the dashboards of their tools and the benefits for DevSecOps team members [W14, W48]. For example, in W48, the speaker discussed how tool dashboards enable collaborative vulnerability management.

\begin{quote}
\textit{So let me just return to the dashboard and point out that it is interactive. (And) I can look at [..] what violations I might have identified in the latest version, a high level summary, a quick bar chart of violations by type, and who has been assigned (to rectify it) [..] We can see that I have one vulnerability assigned to me to go and fix. And then I can indicate it as fixed.} [W48]
\end{quote}

However, speakers noted that many separate reports could be overwhelming with large numbers of tools being used. Therefore, a specific need was identified for tool reports to be compatible with each other (i.e., reports of other tools) [W14]. This feature would enable teams that are using various tools to combine their reports to facilitate comparison and collaboration.

\subsubsection{Configuration}

\textit{\textbf{T4:} Flexibility in configurations and rules.} Since static analysis tools assess the code in its non-operational state, false positives are inevitable. False positives are also a widely reported problem with SAST in the related literature \citep{johnson2013don, christakis2016developers, le2021large}. However, large numbers of false positives in the output of such tools are particularly detrimental for DevOps practitioners who regularly face short delivery milestones. This is because false positives have a very high operational cost, as the developer is required to mute false positives manually and then prioritise the results amidst the pressure to deploy outputs rapidly. For that reason, developer frustrations due to false positives resulting from AST were a source of conflict in this context [W23]. One speaker mentioned that this situation has \textit{led to a lot of finger-pointing and the questioning of why certain (AST tool) results are important} [W23]. Therefore, speakers attempted to provide solutions or workarounds to address it. For example,  proper configuration and setting up of the tool is essential to minimise noise [W4]. Here, speakers emphasised the need for choosing a tool with increased flexibility to configure the relevant rule sets to reduce false positives [W4, W14, W19].

\begin{quote}
\textit{In many cases, we don't want to enable all the rules, and then overburden our developers. So we only enable a few of them.} [W19]
\end{quote}	  

The number of false positives also increases with the number of rules tested in SAST tools. Therefore, flexibility in configuring the rules would allow teams to adjust the scanning function manually, depending on the delivery requirements in DevOps [W19]. More importantly, speakers encouraged developers and security engineers to configure these rule sets in a collaborative manner. 

\subsubsection{Continuous communication}

\textit{\textbf{T5:} Communication platforms.} In the webinar discussions, several speakers highlighted the need for a tool or platform that caters specifically to the communication needs of team members for collaborating in a DevSecOps environment [W8, W19, W20, W46, W47]. Such platforms were seen as beneficial when they were used to deliver the outputs of AST tools across teams. For example, W8 and W46 discussed a workflow where the AST tool is integrated with \textit{Slack}, a popular communications platform \citep{slack_2022}. In this workflow, the discovered vulnerabilities can be efficiently communicated to the developers whenever they are discovered:

\begin{quote}
\textit{You can push these vulnerabilities reported at runtime into Slack, and if your developers are listening to a Slack channel, they will be able to get more information about the vulnerabilities [..]} [W8]
\end{quote}

Challenges related to feedback loops were a specific communications-related issue discussed in our study. Employing a tool or platform for communication was discussed as a solution in the webinars to mitigate this problem [W46]. The use of such tools would enable \textit{continuous communication} between teams, a highly desirable characteristic for this paradigm. In addition, these tools would also ensure another source of records of the communicated details, as comprehensive documentation is typically not maintained in a DevSecOps setting. 

\textit{\textbf{T6:} Team collaboration software.} 
Team collaboration software such as \textit{Confluence} \citep{confluence} or \textit{SharePoint} \citep{sharepoint_2022} were another essential type of tool recommended for CoAST. One use case related to such tools discussed was storing and sharing AST-related process models [W19]. For example, in [W19], the importance of defining the vulnerability discovery and remediation process and sharing via Confluence was discussed. Further, in [W20], collaboration software was used to define and share details related to \textit{engagement models}: 

\begin{quote}
\textit{Getting into a practice of peer reviewing each other's work can help catch overlooked areas and genuinely improve knowledge and awareness. So we start by defining an engagement model. This allows us to have a consistent approach by publishing this on our Confluence page and sharing it internally} [W20]
\end{quote}

In this example, the devised {engagement model} detailed how teams should engage in threat definitions, testing, and collaborating in peer-reviewing this work [W20].

\textit{\textbf{T7:} Automated Bots}. 
Automated Bots (e.g., ChatBots) are software tools that use artificial intelligence or natural language processing techniques to engage with the received input (e.g., messages) \citep{adamopoulou2020chatbots}. Speakers detailed how such Bots can be utilised to enable the \textit{ChatOps} collaboration model (P11):

\begin{quote}
\textit{How to get value from ChatOps?. [..] We've integrated all the tools within Slack. So we have automated bots; every time that the developer raises a PR (pull request), we'll have all of that information fed back to our bots on Slack} [W47] 
\end{quote} 

In ChatOps, one recommendation is to bring the tools (e.g., AST) used in the DevSecOps environment into (typically) a communications platform. However, in that case, the number of messages and artifacts produced by the tools would be difficult to be handled manually. Therefore, bots are introduced to automatically process this information and provide continuous feedback to the developer [W47]. 

\subsubsection{Hybrid testing}
\textit{\textbf{T8:} Interactive Application Security Testing (IAST) tools.} Our analysis shows that tool producers were keen on discussing emerging technologies that addressed long-standing problems in the industry. One such trend was Interactive Application Security Testing (IAST), a hybrid solution combining the features of both SAST and DAST [W6, W8, W11]. IAST functions via software instrumentation to monitor an application as it runs and gathers information about its behaviours. Therefore, this type of tool has the ability to point to true positives (i.e., actual vulnerabilities) in the code. In addition, IAST is able to perform continuous security testing in both custom code and libraries during normal use of the application [W11]. Tool producers also noted that IAST enables a better collaborative workflow for AST:

\begin{quote}
\textit{IAST automatically verifies vulnerabilities to minimise false positives. As a result, you can confidently push seeker reported vulnerabilities to JIRA. Or using Slack, you can send it directly to your developer.} [W8]
\end{quote} 

Due to such reasons, speakers highlighted that IAST suits CoAST, compared with the more traditional AST tools [W8]. 

\subsubsection{Tracking and Correlation}

\textit{\textbf{T9:} Issue or bug tracking systems}. 
AST tools would produce numerous alerts in a CD environment such as DevSecOps. Therefore, vulnerability remediation can be challenging in this setting, particularly when cross-functional team members are involved in the process. Accordingly, speakers proposed using issue tracking systems (e.g., JIRA \citep{jira_2022}) to manage CoAST in this environment [W4, W44]. In [W4],  the speaker emphasised that the activity of traditional security reviews should be replaced by an issue tracking tool-based workflow to enable better collaborative work between the development and security teams:

\begin{quote}
\textit{Developers are generally driven by product development backlogs. They will normally pick work out from our issue tracker system, such as JIRA, (therefore) get your security tooling to be able to inject validated security flaws into (these tracker systems). [..] So it's not something special anymore.} [W4]
\end{quote}

Such AST tool linkage with an issue or bug tracker was recommended in several other webinars [W8, W11, W33, W34]. In this approach, security issues will be seen as another defect or bug to be rectified in the issue tracking system [W4, W11]. Further, the issue tracker links the security engineers' and developers' tools, resulting in a better collaborative workflow [W4]. This workflow would also shorten the loop between when a security issue is detected and the vulnerability remediation [W44]. 

\textit{\textbf{T10:} Security Information and Event Management (SIEM)}. 

In a DevSecOps environment, when teams use different tools for different phases, there is a lack of context in the results, primarily because these systems are disconnected [W48]. Such disconnected tools usually result in alert fatigue and require manual intervention to triage and filter messages [W48]. SIEM tools were recommended by speakers to address this problem [W36, W48]. Such tools aggregate data from multiple sources (e.g., AST tools of different teams), analyse trends and provide actionable inputs to the user [W36].  

\begin{quote}
\textit{It is the onus of the SIEM and its correlation engine to connect all the data produced or alerts produced by these systems to provide enough data to prioritise which vulnerabilities need to be addressed.} [W48]
\end{quote}

The linkage of disparate tools by SIEM eases the burden of both development and security teams [W48]. Furthermore, this tool type ultimately facilitates better collaboration between these teams due to having a single, aggregated and prioritised security information source.

\begin{tcolorbox}[left=1pt, top=1pt, right=1pt, bottom=1pt]
\textbf{RQ3 Summary:} In our study, we extracted several tool capabilities that practitioners should seek to enable a better collaborative workflow in DevSecOps (i.e., T1, T2, T3, T4). We also highlighted the importance of shifting to emerging types of tools that specifically focus on the collaboration needs of modern software development environments such as DevSecOps (e.g., IAST, ChatBots, and Collaboration tools). 
\end{tcolorbox}

\section{Discussion}
In our analysis, we identified several critical drawbacks for CoAST resulting from tool-related issues. In this section, we discuss how these drawbacks can be addressed with improvements to the current tools landscape (i.e., new tool feature requirements). We also discuss key gap areas for future research in this domain based on our analysis. 

\subsection{Feature requirements for collaboration in AST tools}

\begin{figure*} [t]
    \centering
    \includegraphics[scale=.35]{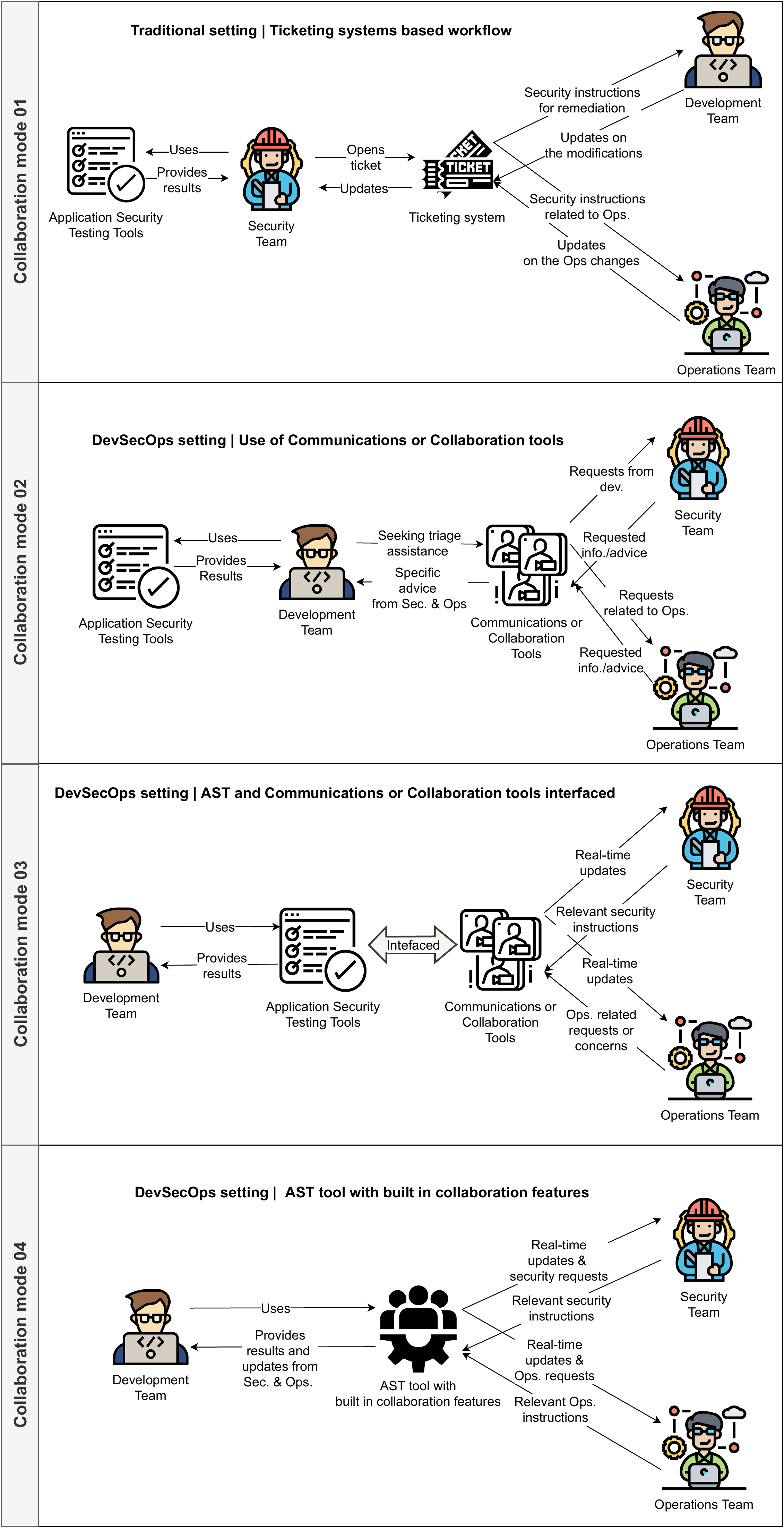}
    \caption{Collaboration modes of DevSecOps team members, AST tools and Communications or collaboration tools}
    \label{fig:modes}
    \vspace{-.3cm}
\end{figure*}

To determine tool features that target CoAST, we established four different \textit{modes} of collaboration between DevSecOps team members, AST tools, and communication or collaboration tools (Fig. \ref{fig:modes}). Here, mode 1 is related to the traditional software development setting, where ticket-based (or email) workflows typically exist. In our study, we presented multiple challenges arising from such a setting for CoAST. Therefore, we propose the below tool feature requirements by analysing these challenges and the currently available solutions.

\subsubsection{Collaboration features built into the AST tools}
The lack of support in AST tools for inter-team collaboration is one of the critical challenges practitioners face when performing CoAST. This problem was reported in the previous literature \citep{tahaei2019survey} and also captured in our analysis (e.g., Ch7). Despite being a significant challenge in this domain, our analysis reveals that it is one area that needs more focus in the industry (e.g., tool producers). Even though our results contain new practices for collaboration (e.g., P11) and emerging AST technologies (e.g., T8), we did not capture any solution that addressed the practitioner need of having collaborative features built into the AST tool itself (Fig. \ref{fig:modes}: mode 4). For example, our analysis showed that sharing code scan outcomes across teams and receiving real-time and continuous feedback in the same AST tool or platform itself would hugely benefit practitioners in a DevSecOps setting. However, we were unable to identify such features from the webinar discussions related to tools or product demonstrations. Therefore, we emphasise that there is a need for features that enable inter-functional team collaboration in AST tools.

There are several ways that the above feature requirement could be technically implemented. For example, tool producers could consider building AST tools that enable different team members to log in to the same platform via the internal network (i.e., Multi-view AST interfaces based on role-based permissions). Studies have proposed similar tool features for collaborative model-driven engineering \citep{franzago2017collaborative}. 

Tool producers could also consider building features for collaboration, such as chat clients or tool capabilities for inter-team code commenting, annotating, tagging, and voting mechanisms in the AST tool. Such features built into the AST tool itself would enable continuous communication and collaboration between team members in the vulnerability remediation process. Further, these features would specifically cater to the developers' requirement of receiving specific actions for vulnerability remediation from the security team (e.g., specific remediation instructions based on code locations). 

The currently available approach to satisfy the above requirements is using different software, such as a collaboration tool, separately to request and receive feedback (Fig. \ref{fig:modes}: mode 2). However, such tool switching in a rapid deployment environment where developers are under delivery pressure is not ideal. Ultimately, this challenge leads to developers being discouraged from engaging in CoAST. 

\begin{tcolorbox}[left=1pt, top=1pt, right=1pt, bottom=1pt]
\textbf{Tool requirement:} There is a need for AST tools that have built-in features for collaboration in the tool for CoAST across teams in DevSecOps environments.
\end{tcolorbox}

\subsubsection{Integration features with communications or collaboration platforms}

In the absence of built-in collaboration features, it is important for AST tools to integrate with other platforms that include such capabilities to promote CoAST. However, our analysis (e.g., Ch8) revealed that most established or legacy AST tools do not have the required capabilities for easy integration (i.e., \textit{a set of disparate, static, disconnected tools} as stated in [W11]). 

Our results indicated that popular collaboration tools provide APIs to support external tool integration \citep{slackapi_2022}. We also presented best practices that recommend integrating AST tools within a communications tool (e.g., ChatOps). However, technically, the AST tool should also have the required interfaces for such integrations. This is clearly a gap area, given the lack of AST tools with such capabilities based on our analysis. Therefore, we recommend that existing AST tool vendors can improve the future releases of their products with this feature (Fig. \ref{fig:modes}: mode 3).

\begin{tcolorbox}[left=1pt, top=1pt, right=1pt, bottom=1pt]
\textbf{Tool requirement:} AST tools require features for integration (e.g., APIs) with communications or collaboration platforms to facilitate CoAST.
\end{tcolorbox}

\subsection{Vulnerability data-centric collaboration and effective visualisation techniques}

Collaboration related to vulnerability data and effective visualisation techniques are areas that need more attention to deal with two key problems in this domain, namely lack of transparency \citep{davis2016effective} and ineffective data sharing practices \citep{colomo2018case, stray2019dependency}. For example, the use of standalone tools by teams resulted in data silos (e.g., Ch9). We captured several practices (e.g., P8, P9) and tool capabilities (e.g., T1, T2, T10) that targeted this specific tool-related challenge. However, several other related issues still need to be adequately addressed.

Speakers proposed using centralised data repositories and implementing data-sharing practices in a DevSecOps environment (P8, P9). However, studies have emphasised the need to share \textit{actionable information} as opposed to large sets of vulnerability data (e.g., scan results) with developers who are under pressure to deliver in this environment \citep{rajapakse2021empirical}. Such actionable information could direct developers to precise actions that are needed to remediate any detected security issues. Other related recommendations include using standard vulnerability scoring systems such as CVSS (i.e., Common Vulnerability Scoring System) \citep{nvd_2022}) to filter and prioritise security data \citep{mell2007complete}. Prioritisation in relation to severity levels is considered an important piece of information when providing vulnerability data to developers in a DevSecOps setting \citep{rajapakse2021empirical}. 

Another underrepresented area in the extracted data is vulnerability data visualisations such as dashboards as a method for collaboration. Most visualisation-related discussions in the webinars were limited to tool functionalities presenting graphs or charts related to the tool's scanning output. However, using visualisations that combine multiple data sources (i.e., tools), perform contextual prioritisation, and provide actionable information are areas that we could not capture in the data. Such visualisation techniques would directly address the reported data sharing and transparency challenges. 

\begin{tcolorbox}[left=1pt, top=1pt, right=1pt, bottom=1pt]
\textbf{Tool requirement:} Shared vulnerability data visualisation techniques (e.g., dashboards) that aggregate data from multiple intra-team sources, analyse and present actionable information for team members is a useful collaboration tool for DevSecOps.
\end{tcolorbox}

\subsection{Socio-technical metrics to measure CoAST}
Software metrics play an important role in obtaining quantitative measurements related to all aspects of software systems. It is also a comparatively well-researched area in software engineering \citep{fenton2000software, ordonez2008state}. However, studies have highlighted the lack of suitable security metrics targeting rapid deployment environments \citep{Rajapakse_Zahedi_Babar_Shen_2022a}. The rapid and continuously changing output makes taking security measurements challenging in DevSecOps. Further, quantitative measurements related to socio-technical matters such as collaboration are also inherently difficult and less researched \citep{ordonez2008state}. 

Despite the above challenges, studies emphasise the importance of establishing suitable metrics to measure collaboration in software development environments. For example, factors such as readiness for collaboration, the quality or costs and benefits of collaboration are highlighted as important to be measured \citep{ordonez2008state}. 

Related to CSE environments, \textit{continuous improvement} is identified as a key continuous practice \citep{fitzgerald2017continuous}. Therefore, suitable metrics should be in place to adopt continuous improvement. However, we noticed that metrics were not a frequently discussed topic in our extracted data.

\textit{Socio-technical metrics} (STMs) are an approach proposed in the literature that focuses on people and their interactions with others in software engineering environments \citep{meneely2014empirical}. An STM represents \textit{the connection between two people in the context of work-related collaboration, which is of a social and technical nature} \citep{trist1951some, meneely2014empirical}. For example, STMs such as \textit{the number of code reviews a source code file has had} have been presented for code review activities \citep{meneely2014empirical}. To address the above-noted metrics-related gap areas, we propose using STMs to study CoAST.

\begin{tcolorbox}[left=1pt, top=1pt, right=1pt, bottom=1pt]
\textbf{Gap area:} Empirically validated socio-technical metrics could be suitable for assessing CoAST in DevSecOps environments.
\end{tcolorbox}

\subsection{Collaboration with AI-enabled agents}
The industry is rapidly moving toward Artificial Intelligence (AI) enabled technologies due to their many benefits. Many AI-enabled tools and services are already available in the DevSecOps industry and related literature (e.g., automated assistance for vulnerability remediation) [W27]. Such technologies are now reducing the role of humans and manual input in these activities. While there were initial perceptions that AI would ultimately replace human workers, authors \citep{hbr2022} emphasise that the way forward is for humans and AI agents to collaborate and enhance each other's strengths. In this scenario, AI-enabled technologies would complement and augment human capabilities, not replace them (\citep{hbr2022} terms this aspect \textit{collaborative intelligence}).

In this evolving technological landscape, DevSecOps team members should now seek to increasingly collaborate with AI agents, as opposed to directly with other teams. For example, developers could pursue the assistance of an AI-enabled agent for vulnerability remediation. Moreover, this agent could communicate remedial measures taken by the developer to the security team (i.e., act as an intermediary). 

AI-enabled chat assistants (e.g., \textit{Conversational AI} \citep{ram2018conversational}, \textit{Interactive conversational systems} \citep{kepuska2018next}) are also rapidly gaining popularity in the industry. Such \textit{smart} chat agents could also be deployed in developer environments for tasks such as providing routine AST knowledge on an on-demand basis. Industry sources also state that AI agents could be useful in bridging collaboration gaps by automatically translating or transcribing conversations and scheduling collaboration sessions \citep{carter_2020}. Such services would be especially helpful if the teams are not co-located.

Despite these benefits, wide adoption of such a working model (i.e., human-AI collaboration for CoAST) has not yet occurred in practice. In the webinars, however, speakers were keen on promoting AI-based tools and services [W27, W31]. In view of this, we suggest that further research is needed to understand the range of technical and socio-technical implications of this emerging area.
 
\begin{tcolorbox}[left=1pt, top=1pt, right=1pt, bottom=1pt]
\textbf{Gap area:} Collaborative intelligence via Human-AI collaboration is an area rapidly gaining practitioner interest. The technical and socio-technical implications of this approach for CoAST need to be further investigated to assess its value for DevSecOps.  
\end{tcolorbox}

\section{Threats to validity}
In this section, we present the threats to validity of our study and actions taken to mitigate them.

\subsection{Data source and generalisability}
Our results are based on the content reported from a selected set of channels. Therefore, there could be threats related to generalising the results, as it is based on these sources. However, the organisers of these webinars or technical talks invited speakers from a range of organisations providing different services and well-known DevOps or DevSecOps experts for their webinars. In addition, many participants from the industry also posed questions at the question and answer sessions. These discussions were also captured in our analysis which was helpful in understanding the challenges faced. Consequently, the results of our study would not be limited to information related to a small set of organisations. 

\subsection{Webinar selection and Quality assessment}
The first author did the filtering task for the selection of webinars and quality assessments. Therefore, the first authors' subjective judgment could be a threat that introduced biases. To mitigate this issue, we strictly followed a pre-defined review protocol. The inclusion, exclusion and quality assessment criteria were reviewed and agreed upon by all authors. We then recorded the webinar lists with the selection decisions based on these criteria using Excel sheets and stored them in shared locations for the review of the remaining authors.

\subsection{Data extraction and Qualitative analysis}
The first author conducted the data extraction and qualitative analysis. Therefore, there could be a threat to validity based on the subjective interpretation of extracted data based on the first authors' viewpoint. To minimise this effect, we carefully followed the recommended steps for thematic synthesis by Cruzes and Dyb\aa \hspace{.5mm} \citep{cruzes2011recommended}. Similar to the webinar selection sheets, the NVivo source files, which contained the extracted data and codes, were stored in a shared location to enable review by the other authors. The coding structure of this study was refined several times in this review. For example, certain themes were expanded, and the analysis was redone for some themes to make the study more focused.

In this article, when presenting our results, we included quotes from webinars for each devised theme for the reader. In addition, we also present summary tables for each RQ that includes the devised themes and key points extracted in our analysis with the webinars that contributed to each key point. By including these details, we aimed to show how the underlying data \textit{fits} each theme. 

\subsection{Marketing material in webinars}
Our data contained discussions from various software development roles, which were beneficial for the aims of our study as we attempted to uncover emerging trends and tool features in the industry. However, certain webinars or parts of the discussions on the webinars (e.g., product demos) included marketing content related to products. Such webinars do not present objective and detailed information (e.g., negative results). To minimise this threat, we included specific exclusion criteria to remove webinars where most of the discussion was centred around marketing a commercial product. 

\section{Conclusion}

In this work, we present a systematisation of knowledge captured from webinars on the collaborative aspect of AST in the DevSecOps paradigm, an area with significantly less focus in the literature. We used thematic analysis on 48 systematically selected webinars from an initial set of 3389 videos from several popular YouTube channels. 

Our results show that DevSecOps practitioners are facing significant challenges that impede continuous collaboration among team members when performing CoAST. We identified nine key challenges categorised into three themes from the thematic analysis of the extracted data. Many challenges were related to people-centric issues, such as role definitions, limitations of AST knowledge and the ownership of security in a DevSecOps setting.

In the selected webinars, speakers were keen on discussing and recommending best practices for collaboration and emerging tools or technologies for DevSecOps. Therefore, we aimed to capture and thematically classify these recommendations from the extracted data. In this effort, we captured eleven best practices in five themes and ten tool types or capabilities in six themes. 

The captured best practices included shifts in the current practices (Shift left and Shift right security), changes to traditional team structures (e.g., Hybrid teams) and roles (e.g., Security and DevOps champions). Speakers also recommended utilising emerging communication practices (e.g., ChatOps), access management strategies (e.g., role-based access) and data-centric collaboration practices (e.g., centralised security data repositories) in this setting.

Regarding tools, speakers strongly recommended considering the integration features (e.g., interfaces) when selecting an AST tool to set up a collaborative environment in DevSecOps. We also captured several existing (e.g., communication platforms and team collaboration software) and emerging tool types (e.g., Automated Bots, IAST) that better facilitate CoAST.

Based on our analysis, we point out the below specific feature requirements for new tool developments and gap areas for further research in this area.

\begin{itemize}

\item There are many communications and collaboration platforms available. However, to facilitate a better collaborative workflow and reduce tool-switching, there is a need for AST tools that have built-in features in the tool itself for CoAST across teams. This requirement is still not adequately fulfilled.

\item Many legacy AST tools are difficult to integrate with other tools or platforms (e.g., communications tools and collaboration platforms). If the tool does not have built-in collaboration capabilities, features that enable easy integration (e.g., API) are required. This is an area where existing tool vendors can improve their future releases. 

\item For vulnerability data-centric collaboration, shared data visualisation techniques (e.g., dashboards) that aggregate data from multiple intra-team sources, contextually prioritise and present actionable information for DevSecOps team members is a useful collaboration tool.

\item There is a lack of empirically validated socio-technical metrics suited to assess collaboration in environments such as DevSecOps that rely on teamwork. Developing such metrics is essential for measuring \textit{continuous improvement} in CoAST, a recommended CSE practice.

\item Finally, collaborative intelligence via Human-AI collaboration is an emerging area rapidly gaining practitioner interest. More research is needed to assess the applicability of this model for CoAST in DevSecOps.

\end{itemize}

\section*{Acknowledgment}
This work has been supported by the Cyber Security Cooperative Research Centre Limited whose activities are partially funded by the Australian Government’s Cooperative Research Centre Programme.

\section*{Declarations}

\subsection*{\textbf{Funding}}
The work has been funded by the Cyber Security Cooperative Research Centre Limited: \url{https://cybersecuritycrc.org.au/}

\subsection*{\textbf{Conflicts of interest}}
The authors have no conflicts of interest to declare.

\subsection*{\textbf{Availability of data}}
The complete transcripts are not released due to copyright policies of YouTube. However, the links of the webinars are included in the appendix.

\subsection*{\textbf{Code availability}}
Not applicable.

\footnotesize
\bibliographystyle{spbasic}

\bibliography{mybibfile.bib} 

\appendix
\section*{Appendix}

\begin{itemize}
\item [W1] DevOpsTV, From Good Code to Great Code  Why Developers Need to Own Application Security, 2017, \url{https://www.youtube.com/watch?v=S9Uno7yyTMI&t=23s}
\item [W2] DevOpsTV, Take Control- Design a Complete DevSecOps Program, 2017, \url{https://www.youtube.com/watch?v=gftkHoPQUIg}
\item [W3] DevSecCon, Building effective DevSecOps teams through role-playing games, 2018, \url{https://www.youtube.com/watch?v=NbAAoXUzIGg}
\item [W4] DevOpsTV, Embrace DevSecOps and Enjoy a Significant Competitive Advantage!, 2018, \url{https://www.youtube.com/watch?v=sv7vVYhxgrs&t=10s}
\item [W5] DevOpsTV, Shift Left Security   The What, Why and How, 2018, \url{https://www.youtube.com/watch?v=I8OSX4Kk97o&t=1s}
\item [W6] DevOpsTV, A Practical Approach to Security Automation, 2019, \url{https://www.youtube.com/watch?v=b-vARs3Wr0o}
\item [W7] DevOpsTV, Shifting Left and Right to Ensure Full Application Security Coverage, 2018, \url{https://www.youtube.com/watch?v=Ya6VX28TNGc}
\item [W8] DevOpsTV, Bridging the Security Testing Gap in Your CI CD Pipeline, 2019, \url{https://www.youtube.com/watch?v=aWWjvy9UpfY&t=1s}
\item [W9] DevOpsTV, Building Resilience into Your DevOps Environment, 2019, \url{https://www.youtube.com/watch?v=-JjERAP9k5Y}
\item [W10] DevSecCon , DevSecOps  Faster Feedback with Security Unit Testing in CI   CD by Eric Johnson  , 2019, \url{https://www.youtube.com/watch?v=Wx7dsbEReiE}
\item [W11] DevOpsTV, Embracing DevSecOps with Embedded Application Security, 2019, \url{https://www.youtube.com/watch?v=JzGBCmIaLAo}
\item [W12] DevSecCon , The New Ways of DevSecOps by James Wickett, 2019, \url{https://www.youtube.com/watch?v=XByV6SBdpYA}
\item [W13] DevOpsTV, 3 Things to Get Right for Successful DevSecOps, 2020, \url{https://www.youtube.com/watch?v=o4_I1GpS62Q}
\item [W14] DevOpsTV, An Open Source DevSecOps Platform for Securing Code \& Dependencies, 2020, \url{https://www.youtube.com/watch?v=MVa9smsCI5Q&t=3s}
\item [W15] DevOpsTV, DevSecOps  Best Practices for Enterprises, 2020, \url{https://www.youtube.com/watch?v=8VO4FZ7YqMI}
\item [W16] Snykcon, DevSecOps for Platform Teams  A Discussion on Making It Easy to Do the Right Thing, 2020, \url{https://www.youtube.com/watch?v=iLvgA8ztBDg}
\item [W17] DevOpsTV, DevSecOps  Closing the Loop from Detection to Remediation, 2020, \url{https://www.youtube.com/watch?v=0ZubO-oty9s&t=24s}
\item [W18] DevOpsTV, Do You Trust Your DevSecOps Pipeline, 2020, \url{https://www.youtube.com/watch?v=X3TMzVKnA1A&t=2s}
\item [W19] DevOpsTV , Don't Let Security Trip You Up- DevOps and Security, 2020, \url{https://www.youtube.com/watch?v=loxRk2tFAzQ}
\item [W20] Snykcon, How to Implement a DevSecOps Culture in a Large Enterprise - People, Processes, Tools, 2020, \url{https://www.youtube.com/watch?v=IJLWMYo1d_A}
\item [W21] DevSecCon , How to implement DevSecOps across the entire organisation, 2020, \url{https://www.youtube.com/watch?v=0IX-h86wfa8}
\item [W22] DevOpsTV, Integrate Security Early and Often For Successful DevSecOps, 2020, \url{https://www.youtube.com/watch?v=hVHbOBsiDv0}
\item [W23] DevOpsTV, Making Security More Efficient for Developers, 2020, \url{https://www.youtube.com/watch?v=4q-dIeI20RQ&t=3s}
\item [W24] DevOpsTV, The DevSecOps Showdown  How to Bridge the Gap Between Security and Developers, 2020, \url{https://www.youtube.com/watch?v=hZ083G_CpLg}
\item [W25] DevSecCon, Trust me, we're doing devsecops - Patrick Debois  , 2020, \url{https://www.youtube.com/watch?v=-hE8o7Yfcko}
\item [W26] DevSecCon, Youve convinced me we have to collaborate but how the hell do we deal with people  Matt Stratton, 2020, \url{https://www.youtube.com/watch?v=B7FTouP3Qas}
\item [W27] DevOpsTV, Are We There Yet  The State of Cloud Native Application Security, 2021, \url{https://www.youtube.com/watch?v=gS2bhGgCGx4}
\item [W28] DevOpsTV, Application Security Moving at the Speed of DevOps  , 2021, \url{https://www.youtube.com/watch?v=Ue05dAZWtHQ}
\item [W29] DevOpsTV, Scaling Governance Compliance and Security Through Pipeline Automation, 2021, \url{https://www.youtube.com/watch?v=uuYSdQ2Y_BI}
\item [W30] DevSecCon, Building a Security Champions program to scale your security team - Simon Maple, 2021, \url{https://www.youtube.com/watch?v=mYqB8zKwt8A}
\item [W31] DevOpsTV, DevOps Trends, Predictions and New Year's Resolutions, 2021, \url{https://www.youtube.com/watch?v=1um2Nj6MwMU}
\item [W32] DevOpsTV, Expect More From Your AppSec Vendor, 2021, \url{https://www.youtube.com/watch?v=ckq7bJT2tTg}
\item [W33] DevOpsTV, Don't Let Third Party Vulnerabilities Run Wild, 2021, \url{https://www.youtube.com/watch?v=33GmhOUTAYM}
\item [W34] OWASP Foundation, AppSec is Dead  Long Live DevSecOps! Matias Madou, 2021, \url{https://www.youtube.com/watch?v=nDQP-kIbOqM}
\item [W35] DevSecCon, How to Influence Developers The Right Way  A Security Team Story - With Kyle Suero and Simon Maple  , 2021, \url{https://www.youtube.com/watch?v=5Tjdtd_V2_A}
\item [W36] DevOpsTV, How to Merge AppSec and DevOps Effectively for the Good of Software  , 2021, \url{https://www.youtube.com/watch?v=DvrMq8fc8DM}
\item [W37] DevOpsTV, The State of DevSecOps  Application Security, 2021, \url{https://www.youtube.com/watch?v=7cCuftPSIU0}
\item [W38] DevOpsTV, Thinking Cloud-Native Application Security  Don’t Let Application Landscapes Complicate Security, 2021, \url{https://www.youtube.com/watch?v=ofFVDXiNNe0}
\item [W39] DevOpsTV, Why Empowering Developers Is a Game Changer for Application Security, 2021, \url{https://www.youtube.com/watch?v=5BJwUseZAUI}
\item [W40] DevOpsTV, DevSecOps  Closing the Security Gap With Developers, 2022, \url{https://www.youtube.com/watch?v=OkGq2r-rDMY}
\item [W41] OWASP Foundation, Developers Struggle with Application Security (and How to Make It Better), 2022, \url{https://www.youtube.com/watch?v=8FQtlbdkXTE}

\item [W42] DevOpsTV, This Year at RSA: Don’t Miss The Conversation on DevSecOps, 2018, \url{https://www.youtube.com/watch?v=t08vNQzSf7s}

\item [W43] DevSecCon, Seven Deadly Saves To Security With Integrations, 2018, \url{https://www.youtube.com/watch?v=1mFOy7z0NY0}

\item [W44] DevOpsTV, Getting Started with Secure DevOps, 2017, \url{https://www.youtube.com/watch?v=D3C318mDsjU}

\item [W45] DevOpsTV, Make Appsec Tools Seamless In Devops Pipelines, 2021, \url{https://www.youtube.com/watch?v=wv8l_WZXeq4}

\item [W46] DevOpsTV, Integrating Security into your Development Pipeline, 2018, \url{https://www.youtube.com/watch?v=_6H_b8C8vJ8}

\item [W47] DevOpsTV, Keep Calm and Secure Your CI CD Pipeline, 2021, \url{https://www.youtube.com/watch?v=yse31jGvNKY}

\item [W48] DevOpsTV, Inserting Security into DevOps Pipelines the Fast Way, 2019, \url{https://www.youtube.com/watch?v=waWS9jfW6Ao}

\end{itemize}

\end{document}